\documentclass[pre,aps,amssymb,amsmath,showpacs,superscriptaddress,twocolumn]{revtex4}

\usepackage{amsmath}
\usepackage{amssymb}
\usepackage{epsfig}
\usepackage{amscd}
\usepackage{graphicx,psfrag,xspace}
\usepackage{float}
\usepackage[normalem]{ulem}

\usepackage{bm}
\usepackage{bbm}

\begin{document}

\title{Neutron fluctuations: the importance of being delayed}
\author{B.~Houchmandzadeh}
\affiliation{CNRS, LIPHY, F-38000 Grenoble, France}
\affiliation{Univ.~Grenoble Alpes, LIPHY, F-38000 Grenoble, France}
\author{E.~Dumonteil}
\author{A.~Mazzolo}
\author{A.~Zoia}
\email{andrea.zoia@cea.fr}
\affiliation{CEA/Saclay, DEN/DANS/DM2S/SERMA/LTSD, 91191 Gif-sur-Yvette, France}

\begin{abstract}
The neutron population in a nuclear reactor is subject to fluctuations in time and in space due to the competition of diffusion by scattering, births by fission events, and deaths by absorptions. As such, fission chains provide a prototype model for the study of spatial clustering phenomena. In order for the reactor to be operated in stationary conditions at the critical point, the population of prompt neutrons instantaneously emitted at fission must be in equilibrium with the much smaller population of delayed neutrons, emitted after a Poissonian time by nuclear decay of the fissioned nuclei. In this work, we will show that the delayed neutrons, although representing a tiny fraction of the total number of neutrons in the reactor, have actually a key impact on the fluctuations, and their contribution is very effective in quenching the spatial clustering.
\end{abstract}


\maketitle

\section{Introduction}
\label{intro}

Many physical and biological systems can be represented in terms of a collection of individuals governed by the competition of the two basic random mechanisms of birth and death. Examples are widespread and encompass neutron multiplication~\cite{pazsit, williams, harris}, nuclear collision cascades~\cite{harris, barucha, athreya}, epidemics and ecology~\cite{bailey, jagers, murray}, bacterial growth~\cite{golding, houchmandzadeh_prl}, and genetics~\cite{lawson, bertoin, sawyer}. Neglecting particle-particle correlations and non-linear effects, the evolution of such systems can be effectively explained by the Galton-Watson model~\cite{harris}. When the death rate is larger than the birth rate, the system is said to be sub-critical: the population size decreases on average, and the ultimate fate is extinction. This occurs for instance for nuclear collision cascades, where charged particles are progressively scattered and absorbed by the medium~\cite{harris, barucha}. When on the contrary the birth rate is larger than the death rate, such as for bacteria reproducing on a Petri dish~\cite{houchmandzadeh_prl}, the system is said to be super-critical. In this case, the population size grows on average. However, because of fluctuations on the number of individuals in the population, a non-trivial finite extinction probability exists for the whole system~\cite{harris}. A super-critical regime is typically found also during the early stages of an epidemic (the so-called `outbreak' phase), where a fast growth of the infected population is observed, until non-linear effects due to the depletion of susceptible individuals ultimately slow down the epidemic~\cite{pnas}. In the intermediate regime, the population stays constant on average, and the system is said to be exactly critical. A prominent example of a system operating at (or close to) the critical point is provided by the self-sustaining fission chains of neutrons in nuclear reactors~\cite{pazsit, williams}.

\begin{figure}[t]
\begin{center}
\includegraphics[scale=0.6]{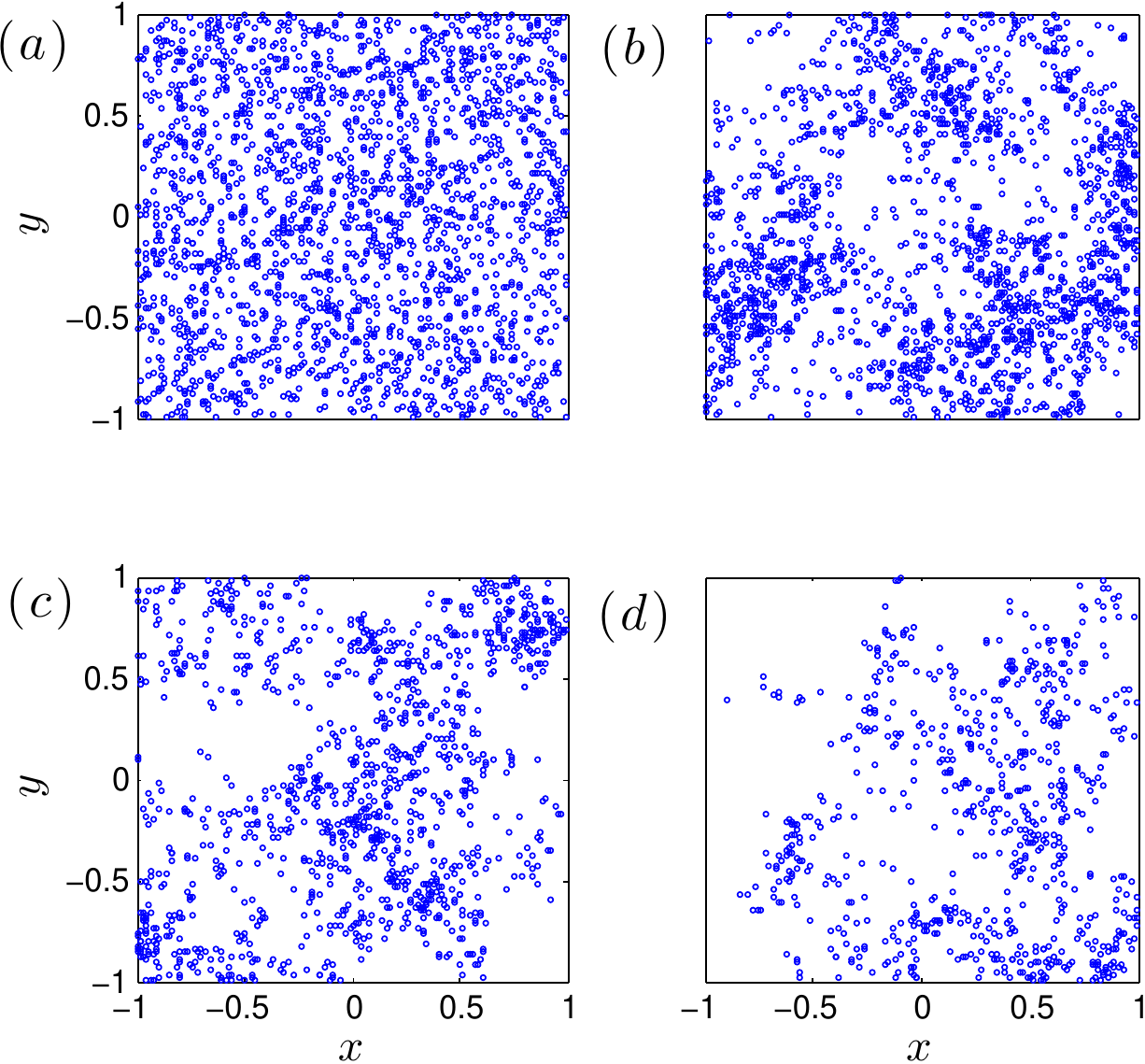}
\end{center}
\caption{(Color online) Monte Carlo simulation of a collection of $10^3$ branching Brownian motions at four successive times: (a) $t=0$, (b) $t=10$, (c) $t=50$, (d) $t=100$. Starting from a uniform spatial distribution at the initial time, the population later shows a wild patchiness due to the interplay between diffusion, reproduction and absorption.}
\label{fig1}
\end{figure}

In most of the models cited above, individuals also interact with the surrounding environment and are typically subject to random displacements~\cite{williams, spatial_eco}. The interplay between the fluctuations stemming from birth-death events and those stemming from diffusion will thus subtly affect the spatial distribution of the particles in such systems~\cite{legall, derrida, derrida_barrier, berestycki_three, majumdar}. Indeed, it has been shown that at and close to the critical point fluctuations due to births and deaths become particularly strong~\cite{harris}: the spatial distribution of the individuals, although uniform at the initial time, may eventually display a wild patchiness (see Fig.~\ref{fig1}), with particles closely packed together and empty spaces nearby~\cite{zhang, meyer, young, houchmandzadeh_prl}. Spatial clustering phenomena have been first identified in connection with mathematical models of ecological communities~\cite{cox, dawson}, and since then have been thoroughly investigated for both infinite and finite collections of individuals~\cite{zhang, meyer, young, cox, dawson, houchmandzadeh_pre_2002, houchmandzadeh_pre_2009, zoia_pre_clustering}.

Non-uniform neutron densities in the reactor fuel elements (which have been named `neutron clustering') might lead to hot spots, and thus represent a most undesirable event with respect to the safe operation of nuclear power plants~\cite{dumonteil_ane, zoia_clustering_jstat}. In view of the relevance of neutron clustering in the context of nuclear reactor physics, in this paper we will investigate the spatial and temporal behaviour of a collection of neutrons undergoing scattering (random displacements), fission (reproduction) and absorption (death). As illustrated in the following, a prototype model of a nuclear reactor can be described in terms of a multi-type branching process involving an equilibrium between neutrons and a second species of particles, the so-called precursors~\cite{harris, williams, pazsit}. Upon fission, a random number of secondary neutrons (called prompt) are emitted almost instantaneously. Precursors are also created by neutrons in very small numbers at fission events, and decay back to neutrons after Poissonian distributed times: such supplementary neutrons stemming from the decay of precursors are called delayed~\cite{bell}.

The central goal of this paper will be to show that the presence of the delayed neutrons, despite their very small amount (typically less than $1$\% of the total number of fission neutrons~\cite{bell}), actually has a quantitative impact on neutron clustering and is very effective in suppressing the fluctuations that are usually observed in birth-and-death systems close to the critical point.

This paper is organized as follows: in Sec.~\ref{reactor} we will present a simple statistical model of fission chains in a nuclear reactor. Then, in Sec.~\ref{statistical_analysis} we will discuss our findings for the physical observable of interest, namely, the average and the second moments of the total neutron and precursor populations. In Sec.~\ref{spatial_behaviour}, we will introduce the two-point correlation functions, so as to extend our analysis to the spatial behaviour of the fluctuations. Conclusions will be finally drawn in Sec.~\ref{conclusions}. Detailed calculations for the physical observables will be provided in series of Appendices.

\section{A prototype model of fission chains in nuclear reactors}
\label{reactor}

\begin{figure}[t]
\begin{center}
\includegraphics[scale=0.6]{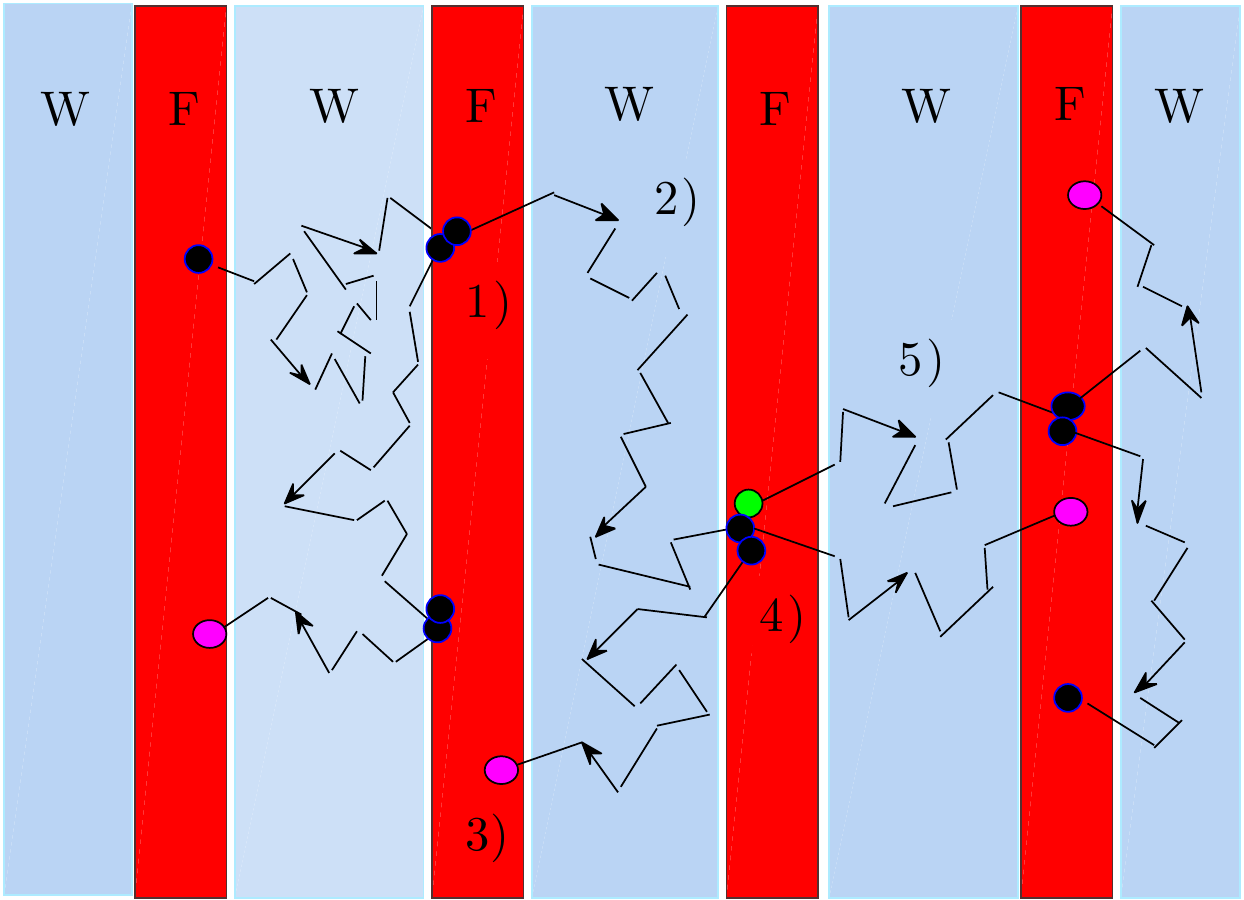}
\end{center}
\caption{(Color online) Simplified sketch of fission chains in a water-moderated reactor. A chain begins with a neutron emitted from a fission event (1) in the fuel (F). The neutron diffuses (2) in water (W) and may later come back to the fuel, where it can either undergo an absorption event (3, marked as a magenta circle), or a new fission event (4). In this latter case, a number of prompt neutrons are emitted instantaneously (marked as black circles). At a later time, following the decay of the excited fissioned nuclei (marked as green circles), a small fraction of delayed neutrons are also emitted and start diffusing in water (5). Both prompt and delayed neutrons contribute to sustaining the fission chains.}
\label{fig2}
\end{figure}

Nuclear reactors are devices aimed at extracting energy from the fission chains induced by neutrons~\cite{bell}. To fix the ideas, here we will roughly sketch the key elements of a water-moderated reactor. The nuclear fuel is basically composed of uranium, arranged in a regular lattice within a cylindrical steel vessel filled with water. A fission chain begins with a neutron emitted at high energy from a fission event on uranium (see Fig.~\ref{fig2}). The neutron enters the surrounding water and progressively slows down towards thermal equilibrium (the so-called moderation phase). Once at thermal energies, the neutron starts diffusing in the water and may eventually re-enter a fuel element of the lattice. Then, the low-energy neutron can either $i)$ be absorbed on the $^{238}U$ isotope of uranium, in which case the chain is terminated; or $ii)$ induce fission on the $^{235}U$ fissile isotope. In this case, the collided nucleus becomes unstable: after a negligible time lapse, it splits into several fragments (typically two) and sets free a variable number of high-energy neutrons (about $2.5$ on average), which are labelled as prompt, and a large amount of energy. The number of fissile nuclei in the reactor core is extremely large, and can be considered constant to a first approximation. The fission fragments are usually left on an excited state and may later decay by a $\beta^-$ nuclear reaction: the energy release on $\beta$-transformation is however in a number of cases sufficiently great to excite the product nucleus to a point where a supplementary high-energy neutron is sent out into the system~\cite{bohr}. Since these extra neutrons are emitted after the decay time of the $\beta^-$ nuclear reactions, they are labelled as delayed (as opposed to prompt). Both prompt and delayed neutrons initiate new fission chains.

The full spatial-dependent behaviour of the neutron population in a nuclear reactor can be fully assessed only by resorting to numerical simulations including a realistic description of the heterogeneous geometry and of the compositions of the core components~\cite{bell, dumonteil_ane}. However, for the purposes of this work we will introduce a prototype model of a nuclear reactor that is simple enough in order for the relevant physical mechanisms to be singled out, and yet retains all the key ingredients of a real system.

The stochastic paths of neutrons within multiplying media are known to follow position- and velocity-dependent random flights~\cite{williams, zoia1, exp_flights}. To begin with, the reactor will be taken to be spatially homogeneous and the random displacements will be approximated by regular $d$-dimensional Brownian motion with a constant diffusion coefficient $D$. The interaction rates of neutrons with matter can be safely assumed to be Poissonian: a neutron is absorbed and disappears at rate $\mu$, and undergoes fission at rate $\alpha$. We will neglect the energy dependence of the probability of fission and capture, and assume that fission can be modelled as a Galton-Watson reproduction process~\cite{harris}: the parent neutron disappears and is replaced by a random number $i$ of identical and independent prompt neutrons, behaving as the parent particle, and a random number $j$ of so-called precursors. The precursors conceptually represent the delayed neutrons being in a `virtual state' before the $\beta^-$ decay of the fission fragments, which sets them free into the system. There exists a joint probability $P_{i,j}$ of generating $i$ neutrons and $j$ precursors at the fission event, the realization $\lbrace i,j \rbrace$ being possibly correlated~\cite{pazsit, williams}. We will denote by $\lambda$ the decay rate of the $\beta^-$ reaction, upon which the precursors disappear to give rise to a delayed neutron.

The mechanism described above formally defines a multi-type branching process~\cite{harris, pazsit}, particle types being neutrons and precursors. Similar behaviours appear also in stochastic biological models of epidemics with dormancy rates, such as for instance for the case of scabies or HIV, where patients with apparent symptoms and patients during incubation would take the roles of neutrons and precursors, respectively~\cite{williams, bailey, jagers, murray}.

\section{Statistical analysis of the total populations}
\label{statistical_analysis}

\begin{figure}[t]
\begin{center}
\includegraphics[scale=0.5]{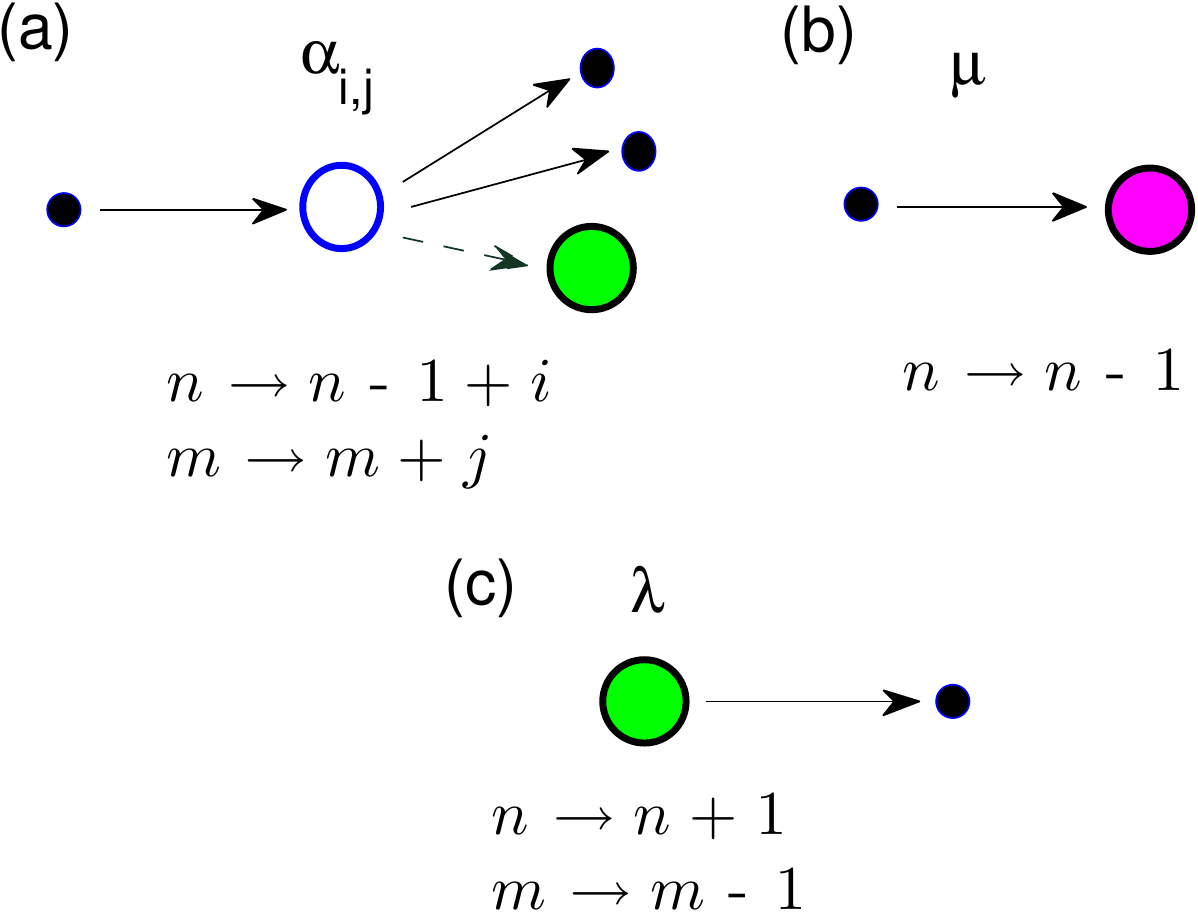}
\end{center}
\caption{(Color online) A scheme of the key events involved in the evolution of the neutron and precursor populations. (a) At fission, a random number of $i$ prompt neutrons and $j$ precursors are created with rate $\alpha_{i,j}$, and the incident neutron is lost.  (b) At absorption, with rate $\mu$, the incident neutron is lost. (c) Upon decay, with rate $\lambda$ a precursor gives rise to a delayed neutron.}
\label{fig3}
\end{figure}

Let us initially consider the evolution of the whole neutron and precursor populations, by ignoring the spatial effects. The system dynamics can be formulated in terms of the transition rates between different discrete states of a two-dimensional Markov chain. Consider a state composed of $n$ neutrons and $m$ precursors at time $t$. Then, the system
\begin{itemize}
\item has a transition $\lbrace n,m \rbrace \to \lbrace n-1, m\rbrace$ with rate $\mu n$,
\item has a transition $\lbrace n,m \rbrace \to \lbrace n-1+i, m+j\rbrace$ with rate $\alpha_{i,j} n= \alpha P_{i,j} n$,
\item has a transition $\lbrace n,m \rbrace \to \lbrace n+1, m-1\rbrace$ with rate $\lambda m$.
\end{itemize}
A scheme is illustrated in Fig.~\ref{fig3}. Following these definitions, the forward master equation for the probability ${\cal P}_t(n,m)$ that at time $t$ the system contains exactly $n$ neutrons and $m$ precursors is given by
\begin{align}
&\frac{\partial}{\partial t} {\cal P}_t(n,m) = -\mu n {\cal P}_t(n,m) - \lambda m {\cal P}_t(n,m) \nonumber \\
& - \sum_{i,j} \alpha_{i,j} n {\cal P}_t(n,m) + \mu (n+1) {\cal P}_t(n+1,m) \nonumber \\
&+ \sum_{i,j} \alpha_{i,j} (n+1-i) {\cal P}_t(n+1-i,m-j)\nonumber \\
& +\lambda (m+1) {\cal P}_t(n-1,m+1).
\label{master_eq}
\end{align}

\subsection{Average number of particles}

By algebraic manipulations, from the master equation we can derive the evolution equations for the moments (see Appendix~\ref{master_moments}). For the average number of particles $\langle n(t) \rangle = \sum_{n,m} n {\cal P}_t(n,m)$ and $\langle m(t)  \rangle = \sum_{n,m} m {\cal P}_t(n,m)$, we get in particular the system
\begin{align}
&\frac{\partial}{\partial t}\langle n(t) \rangle  = (\rho - \alpha \nu_m) \langle n(t) \rangle + \lambda \langle m(t) \rangle \nonumber \\
&\frac{\partial}{\partial t}\langle m(t) \rangle  = \alpha \nu_m \langle n(t) \rangle - \lambda \langle m(t) \rangle,
\label{average_eq}
\end{align}
where we have introduced the average number
\begin{equation}
\nu_n = \sum_{i,j} i P_{i,j}
\end{equation}
of prompt neutrons instantaneously emitted per reproduction event and the average number
\begin{equation}
\nu_m = \sum_{i,j} j P_{i,j}
\end{equation}
of precursors created per reproduction event. The ratio $\nu_m / (\nu_n + \nu_m)$ for water-moderated reactors is about $0.6$\%~\cite{bell}. The quantity $\rho = \alpha(\nu_n + \nu_m -1) - \mu$ physically represents the net reactivity of the system per unit time~\cite{bell}, i.e., the difference between the production rate and the loss rate. For safety reasons, the net reactivity of nuclear reactors is typically weak, in the form of small perturbations around $\rho =0$: this is usually imposed by varying the position of the control elements in the core, which increases or decreases the neutron absorption within the nuclear reactor~\cite{bell}. The system is said to be super-critical if $\rho > 0$, sub-critical if $\rho <0$, and exactly critical if $\rho = 0$.

The evolution of $\langle n(t) \rangle$ and $\langle m(t)\rangle$ is fully determined by assigning the initial conditions $\langle n(0)\rangle=n_0$ and $\langle m(0) \rangle=m_0$. Nuclear reactors are operated at and close to the critical point, so that it is convenient to assume that at time $t=0$ the average neutron and precursor populations are at equilibrium with zero reactivity: this condition is achieved by setting $\partial_t \langle n(t) \rangle\vert_{t=0} = \partial_t \langle m(t) \rangle\vert_{t=0} = 0$, which yields $\alpha \nu_m n_0 = \lambda m_0 $. The quantity $\eta = \lambda / (\alpha \nu_m)$ physically represents the ratio between the rate at which precursors disappear by giving rise to delayed neutrons and the rate at which precursors are created by fission events. In the following, we will always assume that the system is prepared on a zero-reactivity equilibrium configuration at time $t=0$, i.e., $n_0 = \eta m_0$, which implies the initial conditions
\begin{align}
&\langle n(0) \rangle = n_0, \quad{} \langle m(0) \rangle = \frac{n_0}{\eta}.
\label{ic_average}
\end{align}
Actually, one could consider more generally a configuration where precursors are initially absent, and a neutron source is present at time $t=0$. In this case, precursors will be created by fission. If the net reactivity of the system is zero, the number of neutrons will level off to a constant asymptotic value, and so will the number of precursors (see Appendix~\ref{app_mean}). Once equilibrium is attained, one can verify that the ratio between the neutron and precursor population is again $\eta$. In this respect, the main advantage of choosing an initial equilibrium configuration for the two populations is that it allows neglecting the convergence towards the asymptotic equilibrium. Equations~\eqref{average_eq} can be solved exactly (see Appendix~\ref{app_mean}). If the net reactivity is weak, as required above, expanding in small powers of $|\rho|$ yields the asymptotic solutions
\begin{align}
&\langle n(t) \rangle \simeq n_0  \frac{1+ \eta+\epsilon}{1+\eta} e^{\omega t} , \nonumber \\
&\langle m(t) \rangle \simeq m_0 \frac{1+\eta-\eta \epsilon}{1+\eta} e^{\omega t}
\label{average_small_delta}
\end{align}
for long times, where
\begin{align}
& \omega = \frac{\eta}{1+\eta} \rho
\label{asy_delayed_root}
\end{align}
is the characteristic reactor period, and for the sake of convenience we have introduced the rescaled reactivity
\begin{align}
&\epsilon = \frac{\rho}{\alpha \nu_m} \frac{1}{1+\eta}.
\label{asy_reactivity}
\end{align}
The sign of the period $\omega$ depends on the net reactivity: for $\rho >0$, $\omega >0$ and the average populations asymptotically diverge in time; for $\rho <0$, $\omega <0$ and the average populations asymptotically shrink to zero; for $\rho =0$, $\omega =0$ and the populations stay exactly constant in time. For the three cases, the ratio
\begin{align}
& \frac{\langle n(t) \rangle}{\langle m(t) \rangle} \simeq \eta (1+\epsilon)
\label{ratio_n_m_asy}
\end{align}
converges to a constant for long times.

\begin{figure}[t]
\begin{center}
\includegraphics[scale=0.6]{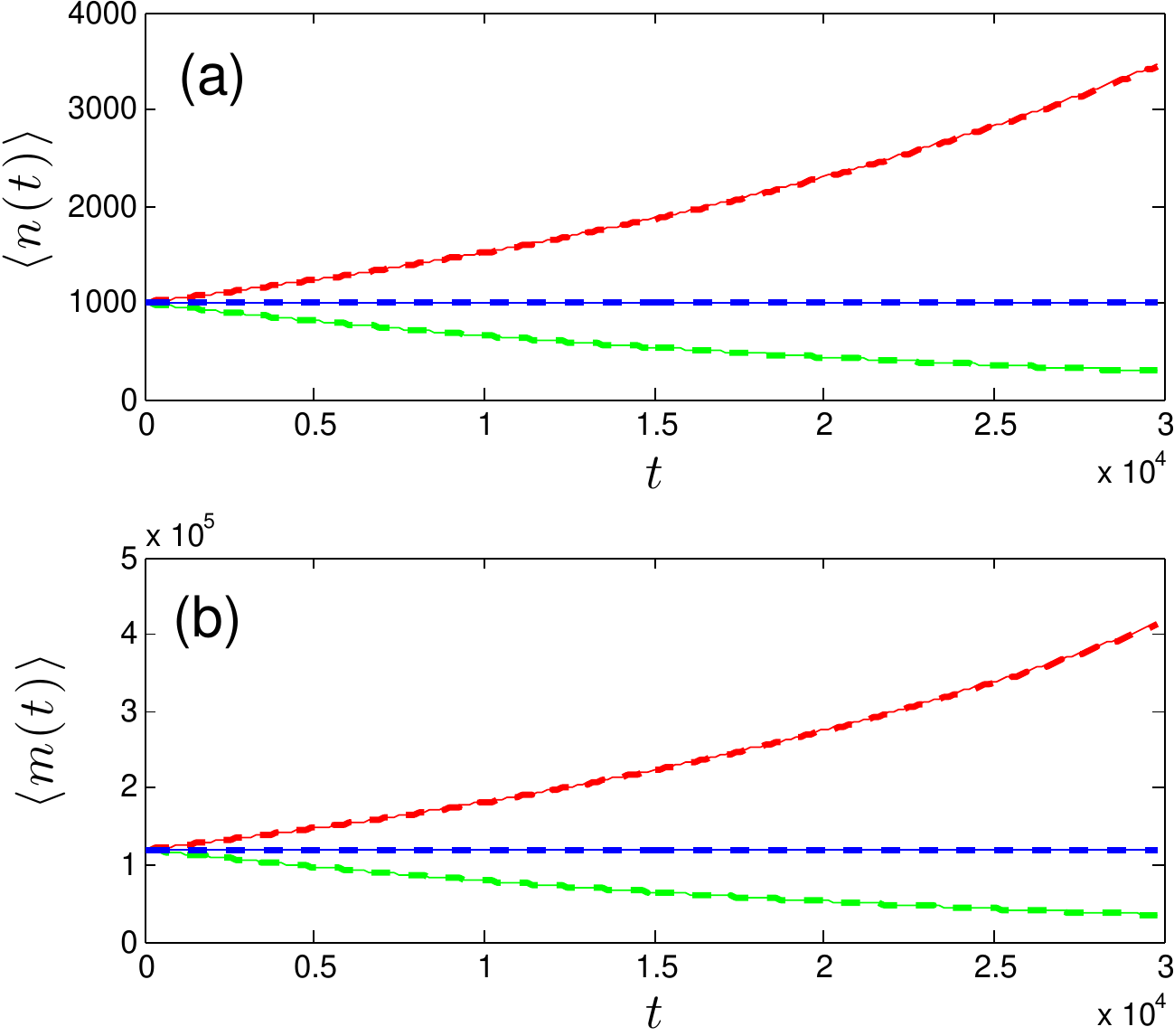}
\end{center}
\caption{(Color online) Evolution of the average neutron population $\langle n(t) \rangle$ (a) and the average precursor population $\langle m(t) \rangle$ (b), starting from a zero-reactivity equilibrium condition. Solid lines are the exact solutions in Eqs.~\eqref{average_eq_exact}, and dashed lines are the asymptotic solutions provided in Eqs.~\eqref{average_small_delta}. The parameters are the following:  $n_0=10^3$, $\eta = 8.333 \times 10^{-3}$, $\alpha \nu_m = 1.2$ and $\lambda = 10^{-2}$. Red (upper) curves correspond to a supercritical reactor with $\rho = 5 \times 10^{-3}$, green (lower) curves correspond to a subcritical reactor with $\rho = -5 \times 10^{-3}$, and blue (central) curves correspond to an exactly critical reactor with $\rho = 0$.}
\label{fig4}
\end{figure}

Equations~\eqref{average_small_delta} are very effective in approximating the exact behaviour of the average neutron and precursor densities in the weak reactivity regime (see Fig.~\ref{fig4} for a numerical example). The accuracy of the approximation increases with decreasing $\eta$. For typical nuclear systems, the coefficient $\eta$ is rather small, about $\eta \simeq 10^{-3}$~\cite{bell}, which follows from the strong separation between the rate at which precursors are created ($\alpha \nu_m$) and the rate at which precursors are converted to delayed neutrons ($\lambda$). This implies that at equilibrium the initial precursor population $m_0 = n_0 /\eta$ is much larger than the initial neutron population $n_0$. Under this assumption, Eqs.~\eqref{average_small_delta} basically say that the average densities have an almost instantaneous jump $n_0 \to n_0 (1+ \eta+\epsilon)/(1+\eta)$ and $m_0 \to m_0 (1+ \eta-\eta\epsilon)/(1+\eta)$, respectively, followed by an exponential growth or decrease (depending on the sign of $\rho$) with an identical period $\omega$. This result is due to the strong separation of the characteristic time scales of the system for small $\eta$ (see Appendix~\ref{app_mean}), and is coherent with the classical results for average observables in reactor physics~\cite{bell}. When $\eta$ is small, precursors have a buffering effect on the evolution of the neutron population: this can be understood by summing up Eqs.~\eqref{average_eq}, which yields
\begin{align}
\frac{\partial}{\partial t}\left[ \langle n(t) \rangle + \langle m(t) \rangle \right] = \rho\langle n(t) \rangle.
\end{align}
Then, from Eq.~\eqref{ratio_n_m_asy} at the leading order we have
\begin{align}
\frac{1+\eta}{\eta} \frac{\partial}{\partial t} \langle n(t) \rangle  = \rho \langle n(t) \rangle,
\end{align}
which implies that reactivity $\rho$ is slowed down by a factor $\eta$ for small values of this parameter.

If the neutron and precursor populations were fully decoupled, and the reactor were to be run based on prompt neutrons alone ($\sum_i P_{i,j} = \delta_{j,0}$, so that $\nu_m =0$, and $\lambda = 0$ for any $m_0$), the net reactivity would be $\rho_p = \alpha(\nu_n -1) - \mu$, and we would have
\begin{align}
&\langle n(t) \rangle_p = n_0 e^{\omega_p t}
\label{average_prompt}
\end{align}
with $\omega_p = \rho_p$~\cite{zoia_pre_clustering}. We have used the subscript $p$ to denote quantities related to purely prompt systems. Since $\nu_m \ll \nu_n$, then $\rho_p \simeq \rho$, whence also $\omega_p \simeq \omega / \eta$. By inspection, we thus have $\langle n(t) \rangle \simeq \langle n(\eta t) \rangle_p$ in the weak reactivity regime. In other words, in the presence of precursors the time at which the neutron population exponentially grows or shrinks is rescaled by a factor $t \to \eta t$, with $\eta \ll 1$. We rediscover here that delayed neutrons, despite their small number, are therefore essential for reactor control thanks to the buffering effect of precursors~\cite{bell}.

\subsection{Equations for the second moments}

It is customary to introduce the normalized and centered second moments, in the form
\begin{align}
&u(t) = \frac{\langle n^2(t)\rangle - \langle n(t)\rangle^2}{\langle n(t)\rangle^2} \\
&v(t) = \frac{\langle n(t)m(t)\rangle - \langle n(t)\rangle \langle m(t)\rangle}{\langle n(t)\rangle \langle m(t)\rangle} \\
&w(t) = \frac{\langle m^2(t)\rangle - \langle m(t)\rangle^2}{\langle m(t)\rangle^2}.
\label{normalized_moments}
\end{align}
The evolution equations for these quantities are derived in Appendix~\ref{app_second_moments}, and read
\begin{align}
\frac{\partial}{\partial t} u(t)  &=- 2 \frac{\lambda}{\chi_t} u(t) + 2 \frac{\lambda}{\chi_t} v(t) \nonumber \\
&+ \frac{1}{\langle n(t) \rangle} \left(\alpha \nu_n^{(2)}+\alpha\nu_m + \frac{\lambda}{\chi_t} -\rho\right),
\label{normalized_u}
\end{align}
\begin{align}
\frac{\partial}{\partial t} v(t)  &= \alpha \nu_m \chi_t u(t) - \left( \frac{\lambda}{\chi_t}  + \alpha \nu_m \chi_t \right) v(t) \nonumber \\
& + \frac{\lambda}{\chi_t} w(t) + \frac{1}{\langle n(t) \rangle}  \left( \alpha \nu_{nm} \chi_t - \alpha \nu_m \chi_t - \lambda \right),
\label{normalized_v}
\end{align}
\begin{align}
\frac{\partial}{\partial t} w(t)  &= 2 \alpha \nu_m \chi_t v(t) -2 \alpha \nu_m \chi_t w(t) \nonumber \\
& + \frac{1}{\langle n(t) \rangle}  \left(\alpha \nu_m^{(2)} \chi^2_t + \alpha \nu_m \chi^2_t + \lambda \chi_t \right),
\label{normalized_w}
\end{align}
where we have defined the factorial moments
\begin{align}
\nu_n^{(2)} = \sum_{i,j} i(i-1) P_{i,j}, \quad{} \nu_m^{(2)} = \sum_{i,j} j(j-1 )P_{i,j}
\end{align}
and the cross-moment
\begin{align}
\nu_{nm} = \sum_{i,j} ij P_{i,j},
\end{align}
and we have set $\chi_t = \langle n(t) \rangle/ \langle m(t) \rangle$. The initial conditions are $u(0) =0$, $v(0) =0$, and $w(0) =0$.

Even though Eqs.~\eqref{normalized_u}-\eqref{normalized_w} can be solved exactly, it is more instructive to focus on their long time behaviour, which is more appropriate for the physical analysis. The asymptotic expansion for small $|\rho|$ and small $\eta $ is detailed in Appendix~\ref{app_uvw}. In particular, under this assumption we can replace $\chi_t \simeq \eta(1+\epsilon)$. By retaining the leading order terms, for long times we have
\begin{align}
u(t) \simeq & \frac{A(1-\epsilon)+\eta(3A+ 2B)}{2 n_0 (1+\eta)(1+\eta - \epsilon)}e^{-\omega t} \nonumber \\
& + \frac{\alpha \nu_m}{n_0} \frac{\eta^2(A+ 2B +C)}{(1+\eta)(1+\eta - \epsilon)} \frac{1-e^{-\omega t} }{\omega},
\label{normalized_u_noncrit}
\end{align}
\begin{align}
v(t) \simeq & \frac{\eta (A+ 2B)}{2 n_0 (1+\eta)(1+\eta - \epsilon)} e^{-\omega t} \nonumber \\
& + \frac{\alpha \nu_m}{n_0} \frac{\eta^2(A+ 2B +C)}{(1+\eta)(1+\eta - \epsilon)} \frac{1-e^{-\omega t} }{\omega},
\label{normalized_v_noncrit}
\end{align}
\begin{align}
w(t) \simeq & \frac{\alpha \nu_m}{n_0} \frac{\eta^2(A+ 2B +C)}{(1+\eta)(1+\eta - \epsilon)} \frac{1-e^{-\omega t} }{\omega},
\label{normalized_w_noncrit}
\end{align}
where the coefficients read
\begin{align}
& A = \frac{\nu_n^{(2)}}{\nu_m} \left( 1-\frac{\epsilon}{1+\eta}\right) + 2\left( 1-2\frac{\epsilon}{1+\eta}\right),
\end{align}
$B = \nu_{nm}/ \nu_m -2$, and $C = \nu_m^{(2)}/\nu_m+2$.

\begin{figure}[t]
\begin{center}
\includegraphics[scale=0.6]{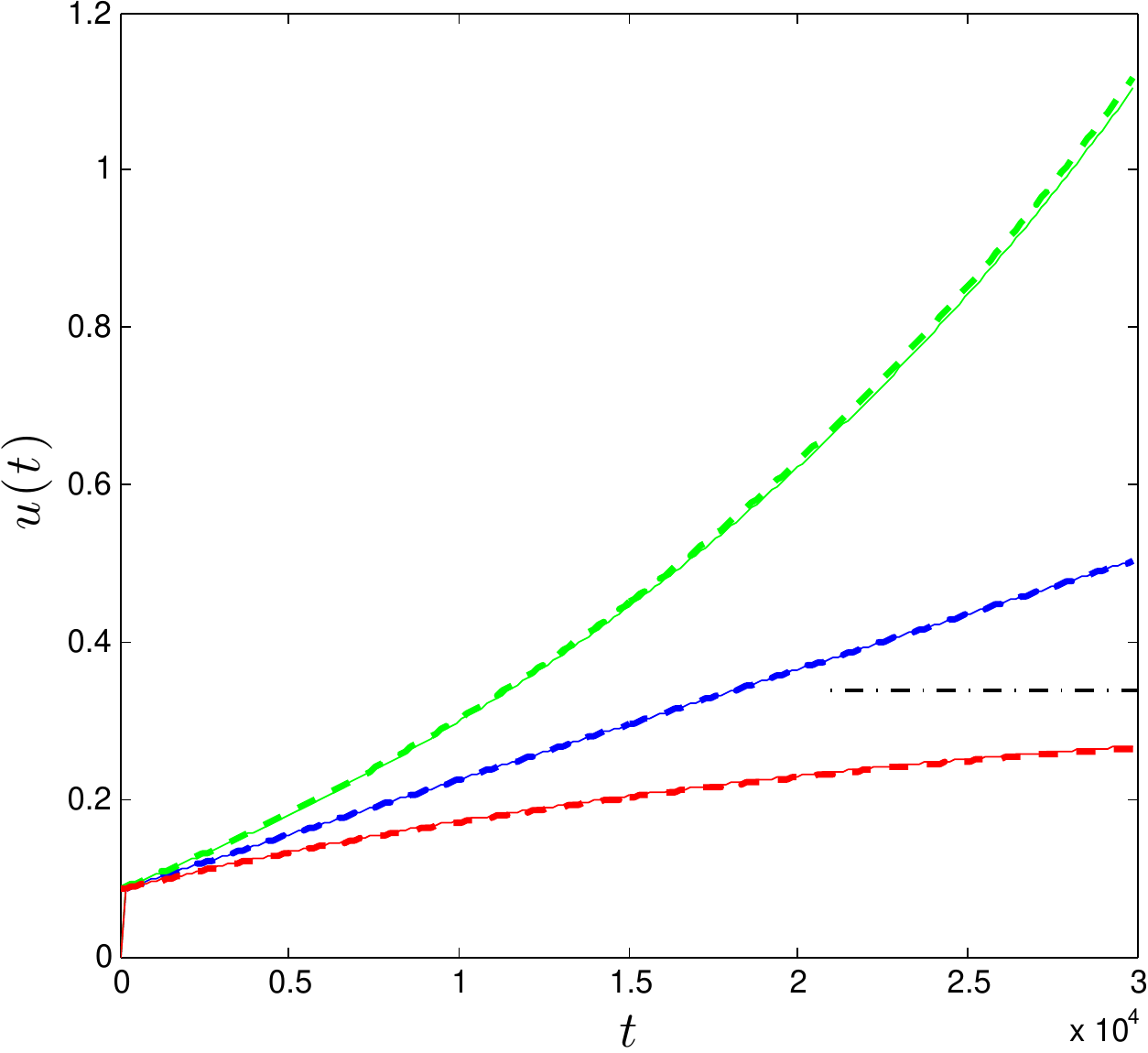}
\end{center}
\caption{(Color online) Evolution of the normalized and centered second moment of the neutron population $u(t)$, starting from a zero-reactivity equilibrium condition. Solid lines are the numerical solutions of the exact Eq.~\eqref{normalized_u}, and dashed lines are the asymptotic solutions provided in Eqs.~\eqref{normalized_u_noncrit} and~\eqref{normalized_u_crit}. The parameters are the following: $n_0=10^3$, $\eta = 8.333 \times 10^{-3}$, $\alpha \nu_m = 1.2$, $\lambda = 10^{-2}$, $\nu_n^{(2)} = 2$, $\nu_{nm} = 2.4 \times 10^{-2}$ and $\nu_m^{(2)} = 4\times 10^{-3}$. Red (lower) curves correspond to a supercritical reactor with $\rho = 5 \times 10^{-3}$, green (upper) curves correspond to a subcritical reactor with $\rho = -5 \times 10^{-3}$, and blue (central) curves correspond to an exactly critical reactor with $\rho = 0$. The dotted-dashed black line corresponds to the asymptotic value $u_\infty$ expected for the supercritical configuration, as given in Eq.~\eqref{normalized_u_asy}.}
\label{fig5}
\end{figure}

\begin{figure}[t]
\begin{center}
\includegraphics[scale=0.6]{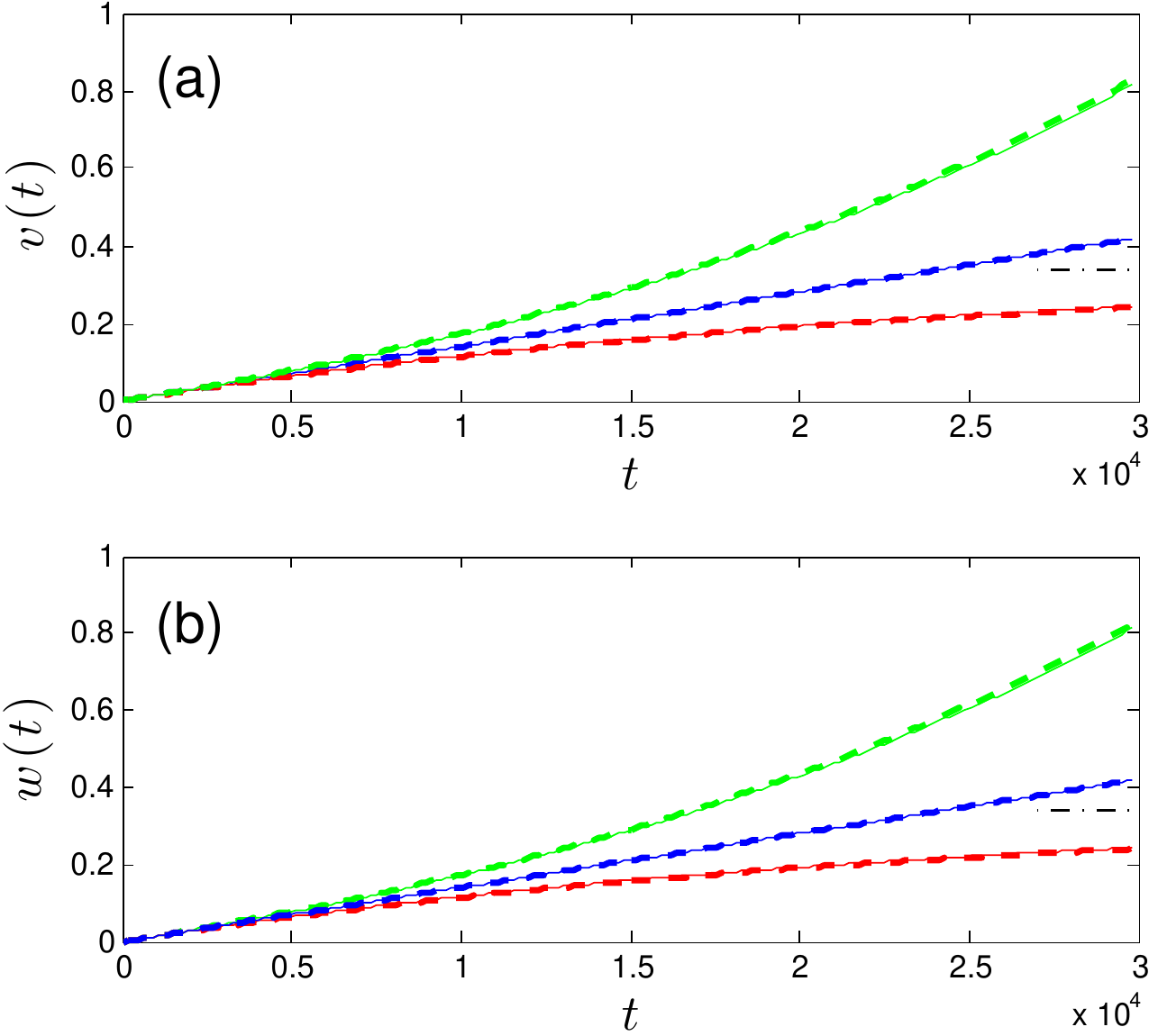}
\end{center}
\caption{(Color online) Evolution of the normalized and centered second moments $v(t)$ (a) and $w(t)$ (b), starting from a zero-reactivity equilibrium condition. Solid lines are the numerical solutions of the exact Eqs.~\eqref{normalized_v} and~\eqref{normalized_w}, respectively, and dashed lines are the asymptotic solutions provided in Eq.~\eqref{normalized_v_noncrit} (Eq.~\eqref{normalized_v_crit} for the critical case) and Eq.~\eqref{normalized_v_noncrit} (Eq.~\eqref{normalized_v_crit} for the critical case), respectively, for small reactivities and $\eta \ll 1$. The parameters are the following: $n_0=10^3$, $\eta = 8.333 \times 10^{-3}$, $\alpha \nu_m = 1.2$, $\lambda = 10^{-2}$, $\nu_n^{(2)} = 2$, $\nu_{nm} = 2.4 \times 10^{-2}$ and $\nu_m^{(2)} = 4\times 10^{-3}$. Red (lower) curves correspond to a supercritical reactor with $\rho = 5 \times 10^{-3}$, green (upper) curves correspond to a subcritical reactor with $\rho = -5 \times 10^{-3}$, and blue (central) curves correspond to an exactly critical reactor with $\rho = 0$. The dotted-dashed black line corresponds to the asymptotic value $v_\infty = w_\infty$ expected for the supercritical configuration, as given in Eq.~\eqref{normalized_u_asy}.}
\label{fig6}
\end{figure}

The asymptotic expressions in Eqs.~\eqref{normalized_u_noncrit},~\eqref{normalized_v_noncrit} and~\eqref{normalized_w_noncrit} are compared to the numerical solutions of the exact Eqs.~\eqref{normalized_u},~\eqref{normalized_v} and~\eqref{normalized_w} in Figs.~\ref{fig5} and~\ref{fig6}. The agreement of the asymptotic to the exact solutions in the weak reactivity regime is remarkable also for the second moments of the populations. When the reactor is subcritical, the average neutron and precursor populations decrease exponentially fast, so that $u(t)$, $v(t)$ and $w(t)$ diverge exponentially fast as $\sim \exp(|\omega| t)$. At some point, $u(t)$, $v(t)$ and $w(t)$ will become larger than $1$, and the fluctuations will completely overrule the average behaviour of the individuals. On the contrary, when the reactor is supercritical the average neutron and precursor populations grow unbounded exponentially fast, and $u(t)$, $v(t)$ and $w(t)$ saturate to the asymptotic value
\begin{align}
& u_\infty = v_\infty = w_\infty \simeq \frac{\alpha \nu_m}{n_0} \frac{\eta^2(A+ 2B +C)}{\omega(1+\eta)(1+\eta - \epsilon)}.
\label{normalized_u_asy}
\end{align}
In this regime, the fluctuations may still be large when the initial neutron population $n_0$ is small. The case of an exactly critical system can be obtained from the previous equations by taking the limit for $\omega \to 0$, with $\epsilon=0$. We thus obtain
\begin{align}
& u(t) \simeq \frac{A+\eta(3A+ 2B)}{2 n_0 (1+\eta)^2} + \frac{\alpha \nu_m}{n_0} \frac{\eta^2(A+ 2B +C) }{(1+\eta)^2} t,
\label{normalized_u_crit}
\end{align}
\begin{align}
& v(t) \simeq \frac{\eta (A+ 2B)}{2 n_0 (1+\eta)^2} + \frac{\alpha \nu_m}{n_0} \frac{\eta^2(A+ 2B +C)}{(1+\eta)^2} t,
\label{normalized_v_crit}
\end{align}
\begin{align}
& w(t) \simeq\frac{\alpha \nu_m}{n_0} \frac{\eta^2(A+ 2B +C)}{(1+\eta)^2} t.
\label{normalized_w_crit}
\end{align}
In the critical regime, the average populations stay constant, but the moments $u(t)$, $v(t)$ and $w(t)$ diverge linearly in time and will ultimately cross the threshold at one: this stems from the individuals being (almost surely) doomed to extinction~\cite{williams}. The typical extinction time $\tau_E$ for the neutron population can be determined by imposing $u(\tau_E) \simeq 1$, which yields
\begin{equation}
\tau_E =  \frac{n_0}{\alpha \nu_m (A+2B+C) \eta^2}
\end{equation}
by neglecting sub-leading order terms.

If the reactor were to be operated with prompt neutrons alone, we would have
\begin{align}
& u_p(t) = \frac{\alpha \nu_n^{(2)}}{n_0} \frac{1-e^{-\omega_p t} }{\omega_p},
\label{normalized_u_noncrit_prompt}
\end{align}
for $\omega_p \neq 0$, and 
\begin{equation}
u_p(t) = \frac{\alpha \nu_n^{(2)}}{n_0} t
\end{equation}
in the critical regime~\cite{zoia_pre_clustering}. By direct inspection, observing that the term $\nu_m (A+2B+C)$ is dominated by $\nu_n^{(2)}$ since $\sum_i P_{i,j} \ll \sum_j P_{i,j}$ for $j \ge 1$, we have
\begin{equation}
u(t) \simeq \eta u_p( \eta t).
\end{equation}
In other words, in the presence of precursors, the normalized and centered second moment of the neutron population has a much slower evolution in time ($t \to \eta t$, similarly as for the case of the average number of particles), and its amplitude is further rescaled by a factor $\eta$. As for the extinction time, $\tau_E^p \simeq n_0 /(\alpha \nu_n^{(2)})$, and we would have $\tau_E \simeq \tau_E^p / \eta^2$. In this respect, precursors are extremely effective in quenching the neutron fluctuations at the scale of the global population.

\section{Spatial behaviour of the populations}
\label{spatial_behaviour}

We would like now to address the spatial behaviour of neutrons and precursors. For the sake of simplicity, let us initially consider a one-dimensional domain partitioned into cells of size $\ell$, the cell of index $k$ containing $n_k$ neutrons and $m_k$ precursors. The full state of the particles will be provided by the vectors $({\bf n}, {\bf m})$, where ${\bf n} = \left\lbrace ..., n_k ,...\right\rbrace$ and ${\bf m} = \left\lbrace ..., m_k ,...\right\rbrace$. In order to manipulate a modified state where a particle has been added or removed from the site $k$ with respect to ${\bf n}$, it is convenient to resort to the formalism proposed in~\cite{houchmandzadeh_pre_2002, paessens, durang, tauber}: we will introduce the creation and annihilation operators $a^\dag_k$ and $a_k$, whose action on ${\bf n}$ yields $a^\dag_k{\bf n} = (...,n_{k-1},n_k + 1,n_{k+1},...)$ and $a_k{\bf n} = (...,n_{k-1},n_k - 1,n_{k+1},...)$, respectively, and the operators $b^\dag_k$ and $b_k$ that have identical action on ${\bf m}$. Assume that the reactor is in state $({\bf n}, {\bf m})$ at time $t$. Then, the system
\begin{itemize}
\item has a transition $\lbrace {\bf n}, {\bf m} \rbrace \to \lbrace a_k {\bf n}, {\bf m} \rbrace$ with rate $\mu _kn$,
\item has a transition $\lbrace {\bf n}, {\bf m} \rbrace \to \lbrace (a^\dag_k)^i a_k {\bf n}, (b^\dag_k)^j {\bf m} \rbrace$ with rate $\alpha_{i,j} n= \alpha P_{i,j} n_k$,
\item has a transition $\lbrace {\bf n}, {\bf m} \rbrace \to \lbrace a^\dag_k {\bf n}, b_k {\bf m}  \rbrace$ with rate $\lambda m_k$,
\item has a transition $\lbrace {\bf n}, {\bf m} \rbrace \to \lbrace a^\dag_{k\pm1} {\bf n}, {\bf m}  \rbrace$ with rate $\gamma n_k$,
\end{itemize}
where $\gamma$ is the diffusion rate of neutrons from neighbouring cells $k \pm 1$. Following these definitions, the forward master equation for the probability ${\cal P}_t({\bf n}, {\bf m})$ that at time $t$ the system is in state $({\bf n}, {\bf m})$ obeys
\begin{align}
&\frac{\partial}{\partial t} {\cal P}_t({\bf n}, {\bf m}) = \sum_k \Big[ - \sum_{i,j} \alpha_{i,j} n_k {\cal P}_t({\bf n}, {\bf m}) \nonumber \\
& + \sum_{i,j} \alpha_{i,j} (n_k+1-i) {\cal P}_t((a^\dag_k)^i a_k {\bf n}, (b^\dag_k)^j {\bf m}) \nonumber \\
& -\mu n_k {\cal P}_t({\bf n}, {\bf m}) + \mu (n_k+1) {\cal P}_t(a_k^\dag{\bf n}, {\bf m})\nonumber \\
& - \lambda m_k {\cal P}_t({\bf n}, {\bf m}) +\lambda (m_k+1) {\cal P}_t(a_k{\bf n}, b_k^\dag{\bf m}) \nonumber \\
& - 2\gamma n_k{\cal P}_t({\bf n}, {\bf m}) +\gamma (n_{k+1}+1){\cal P}_t(a_ka^\dag_{k+1}{\bf n}, {\bf m}) \nonumber \\
&  +\gamma (n_{k-1}+1){\cal P}_t(a_ka^\dag_{k-1}{\bf n}, {\bf m})\Big].
\label{master_eq_space}
\end{align}
Generally speaking, the solutions of Eq.~\eqref{master_eq_space} could be sought by resorting to a field-theoretical approach~\cite{tauber}. However, thanks to the master equation being linear, the equations for the spatial moments of the population can again be obtained by simpler algebraic manipulations of Eq.~\eqref{master_eq_space} (see Appendix~\ref{master_moments}).

\subsection{Average particle densities}

For the average number of particles in a cell $k$, namely,
\begin{align}
&\langle n_k(t) \rangle = \sum_{{\bf n}, {\bf m}} n_k {\cal P}_t({\bf n}, {\bf m}) \nonumber \\
&\langle m_k(t)  \rangle = \sum_{{\bf n}, {\bf m}} m_k {\cal P}_t({\bf n}, {\bf m}),
\end{align}
we get in particular the system
\begin{align}
&\frac{\partial}{\partial t}\langle n_k(t) \rangle  = (\gamma \Delta+ \rho - \alpha \nu_m) \langle n_k(t) \rangle + \lambda \langle m_k(t)\rangle,\nonumber \\
&\frac{\partial}{\partial t}\langle m_k(t) \rangle  = \alpha \nu_m \langle n_k(t) \rangle - \lambda \langle m_k(t) \rangle,
\label{average_eq_space}
\end{align}
where we have used the shorthand notation $\Delta f_k = f_{k+1}- 2 f_k + f_{k-1}$ for the discrete Laplacian operator. We can then define the average densities of neutrons and precursors by taking the continuum limit
\begin{align}
& {\mathcal N}(x,t) = \lim_{\ell \to 0} \frac{\langle n_k(t) \rangle}{\ell}, \quad {\mathcal M}(x,t) = \lim_{\ell \to 0} \frac{\langle m_k(t)\rangle}{\ell},
\label{normalized_concentrations}
\end{align}
where $x = k \ell$. By replacing these definitions in the previous equations, the average densities satisfy
\begin{align}
&\frac{\partial}{\partial t}{\mathcal N}(x,t)  =  (D \nabla^2 + \rho - \alpha \nu_m){\mathcal N}(x,t)+ \lambda {\mathcal M}(x,t) ,\nonumber \\
&\frac{\partial}{\partial t}{\mathcal M}(x,t)  = \alpha \nu_m {\mathcal N}(x,t) - \lambda {\mathcal M}(x,t),
\label{average_conc_eq}
\end{align}
where we have used the Taylor expansion $ \langle \Delta n_k(t) \rangle \simeq \ell^2 \nabla^2 {\mathcal N}(x,t)$, and $D = \lim_{\ell \to 0} \gamma \ell^2$ is the diffusion coefficient of the neutrons.

Imposing as above the zero-reactivity equilibrium initial conditions leads to ${\mathcal N}_0 = {\mathcal N}(x,0) = \eta {\mathcal M}(x,0)$, which means that the spatial profile of the neutron and precursor concentrations will be flat. By inspection of Eq.~\eqref{average_conc_eq}, it is apparent that starting from this initial condition the concentrations will stay flat, and that their amplitude will follow the same time behaviour as the average total populations $\langle n(t) \rangle $ and $\langle m(t) \rangle $ in Eq.~\eqref{average_eq_exact}. In particular, for weak reactivities at long times we have
\begin{align}
& {\mathcal N}(x,t) = {\mathcal N}(t) \simeq {\mathcal N}_0 \frac{1 + \eta + \epsilon}{1+\eta} e^{\omega t}
\end{align}
and the ratio ${\mathcal N}(x,t)/ {\mathcal M}(x,t)$ asymptotically converges again to the constant $\eta(1+\epsilon)$. Observe that the only dimension-dependent term in Eq.~\eqref{average_conc_eq} is the spatial derivative $\nabla^2$, so that the evolution equations for the concentration would be left almost unchanged in a $d$-dimensional infinite space $\mathbb R^d$, provided that we replace $x$ with ${\bf r}$ and $\nabla^2$ with the $d$-dimensional Laplacian $\nabla^2_{d}$.

Finally, by analogy with the case of the average neutron population, in the weak reactivity regime we have ${\mathcal N}(t) \simeq {\mathcal N}_p(\eta t)$, where ${\mathcal N}_p(\eta t)$ is the average neutron density for a reactor that were to be run based on prompt neutrons alone.

\subsection{Spatial correlation functions}

We will define the spatial correlation functions
\begin{align}
&\langle n_k n_{k+j} \rangle = \sum_{{\bf n}, {\bf m}} n_k n_{k+j}  {\cal P}_t({\bf n}, {\bf m}) \nonumber \\
&\langle m_k m_{k+j} \rangle = \sum_{{\bf n}, {\bf m}} m_k m_{k+j} {\cal P}_t({\bf n}, {\bf m})
\end{align}
and the cross-correlations
\begin{equation}
\langle n_{k+j} m_k \rangle = \sum_{{\bf n}, {\bf m}} n_{k+j} m_k {\cal P}_t({\bf n}, {\bf m}).
\end{equation}
We have dropped the explicit time dependence for the sake of conciseness. Assuming that the initial particle concentrations are spatially flat allows applying a translational symmetry to the system (in particular, for the averages we have $\langle n_k \rangle = \langle n_{k+j} \rangle$ and $\langle m_k \rangle = \langle m_{k+j} \rangle$ $\forall k,j$). It is then convenient to introduce the normalized and centered moments
\begin{align}
& u_j(t) = \frac{\langle n_k n_{k+j} \rangle}{\langle n_k \rangle^2} -1 - \frac{\delta_{j,0}}{\langle n_k \rangle} \nonumber \\
& v_j(t) = \frac{\langle n_{k+j} m_k \rangle}{\langle n_k \rangle \langle m_k \rangle} -1 \nonumber \\
& w_j(t) = \frac{\langle m_k m_{k+j} \rangle}{\langle m_k \rangle^2} -1 - \frac{\delta_{j,0}}{\langle m_k \rangle},
\label{normalized_uvw}
\end{align}
which only depend on the relative distance $|j|$ between site $k$ and $k+j$~\cite{houchmandzadeh_pre_2002}. The Kronecker delta term $\delta_{i,j}$ expresses the contribution of self-correlations. The evolution equations for the spatial correlations are provided in Appendix~\ref{app_spatial_second_moments}, and read
\begin{align}
\frac{\partial}{\partial t} u_j &= 2 (\gamma \Delta -\frac{\lambda}{\chi_t} ) u_j + 2 \frac{\lambda}{\chi_t} v_j+\alpha \nu_n^{(2)} \frac{\delta_{j,0}}{\langle n_k \rangle}, \\
\frac{\partial}{\partial t} v_j &= \alpha \nu_m u_j + (\gamma \Delta - \alpha \nu_m\chi_t - \frac{\lambda}{\chi_t} )v_j \nonumber \\
& +\frac{\lambda}{\chi_t} w_j + \alpha \nu_{nm}\chi_t \frac{\delta_{j,0}}{\langle n_k \rangle},\\
\frac{\partial}{\partial t} w_j & = 2 \alpha \nu_m \chi_t v_j  - 2 \alpha \nu_m \chi_t w_j +\alpha \nu_m^{(2)} \chi^2_t \frac{\delta_{j,0}}{\langle n_k \rangle},
\label{uvw_eq_space_simple}
\end{align}
where we have used $\langle n_k \rangle / \langle m_k \rangle = \chi_t$. By taking again the continuum limit $\ell \to 0$, with $r = \ell |j|$ and $\gamma \Delta f_j \simeq D \nabla^2 f(r) $, we finally obtain the evolution equations for the correlations $u(r,t) = \lim_{\ell \to 0} u_j(t)$, $v(r,t) = \lim_{\ell \to 0} v_j(t)$, and $w(r,t) = \lim_{\ell \to 0} w_j(t)$, namely
\begin{align}
\frac{\partial}{\partial t} u(r,t)  &= 2( D \nabla^2 -\frac{\lambda}{\chi_t} )u(r,t) + 2 \frac{\lambda}{\chi_t} v(r,t) \nonumber \\
& + \alpha \nu_n^{(2)} \frac{\delta(r)}{{\mathcal N}(t)} ,
\label{normalized_u_corr}
\end{align}
\begin{align}
\frac{\partial}{\partial t} v(r,t)  &= (D \nabla^2  -  \frac{\lambda}{\chi_t} - \alpha \nu_m \chi_t ) v(r,t) +\alpha \nu_m \chi_t u(r,t) \nonumber \\
&  + \frac{\lambda}{\chi_t} w(r,t) + \alpha \nu_{nm} \chi_t \frac{\delta(r)}{{\mathcal N}(t)} ,
\label{normalized_v_corr}
\end{align}
\begin{align}
\frac{\partial}{\partial t} w(r,t)  & = 2 \alpha \nu_m \chi_t v(r,t) -2 \alpha \nu_m \chi_t w(r,t) \nonumber \\
& + \alpha \nu_m^{(2)} \chi^2_t \frac{\delta(r)}{{\mathcal N}(t)} .
\label{normalized_w_corr}
\end{align}
These equations hold true in any dimension $d$, provided that $\nabla^2$ is replaced by the $d$-dimensional Laplacian $\nabla^2_d$.

The long time and long distance expansion of the previous equations for small $|\rho|$ and small $\eta$ is discussed in Appendix~\ref{app_uvw_spatial} for a $d$-dimensional domain. By retaining the leading order terms, in this regime we obtain
\begin{align}
u(r,t) &\simeq \frac{\alpha \nu_m A'(1-\epsilon) }{2 {\mathcal N}_0 (1+\eta)(1+\eta - \epsilon)} H(r,t)\nonumber \\
& + \frac{\alpha \nu_m}{{\mathcal N}_0} \frac{\eta^2(A'+ 2B' +C')}{(1+\eta)(1+\eta - \epsilon)} F(r,t),
\label{normalized_u_space_real}
\end{align}
\begin{align}
 v(r,t) &\simeq \frac{\alpha \nu_m \eta (A'+ 2B') }{2 {\mathcal N}_0 (1+\eta)(1+\eta - \epsilon)} H(r,t) \nonumber \\
& + \frac{\alpha \nu_m}{{\mathcal N}_0} \frac{\eta^2(A'+ 2B' +C')}{(1+\eta)(1+\eta - \epsilon)} F(r,t),
\label{normalized_v_space_real}
\end{align}
\begin{align}
w(r,t) & \simeq \frac{\alpha \nu_m}{{\mathcal N}_0} \frac{\eta^2(A'+ 2B' +C')}{(1+\eta)(1+\eta - \epsilon)} F(r,t),
\label{normalized_w_space_real}
\end{align}
where we have defined the quantities
\begin{align}
& F(r,t) = e^{-\omega t} \int_0^t dt' e^{\omega t'} G(r,t'),
\end{align}
where $G(r,t)$ is the Gaussian function
\begin{align}
& G(r,t) = \frac{\exp\left(-\frac{r^2}{8 \eta {\mathcal D} t} \right) }{\left( 8 \pi \eta {\mathcal D} t\right)^{d/2}},
\end{align}
and
\begin{align}
& H(r,t) = \frac{e^{-\omega t}}{\alpha \nu_m} \left( \frac{\alpha \nu_m}{{\mathcal D}}\right)^{\frac{d+2}{4}} \frac{K_{d/2-1}\left(r\sqrt{\alpha \nu_m / {\mathcal D}}\right)}{(2 \pi)^{d/2} r^{d/2-1}},
\end{align}
$K_a(z)$ being the modified Bessel function of the second kind~\cite{erdelyi}. The parameters read
\begin{equation}
A' = \frac{\nu_n^{(2)}}{\nu_m} ,
\end{equation}
$B' = \nu_{nm}/ \nu_m$ and $C' = \nu_m^{(2)}/\nu_m$.

\begin{figure}[t]
\begin{center}
\includegraphics[scale=0.6]{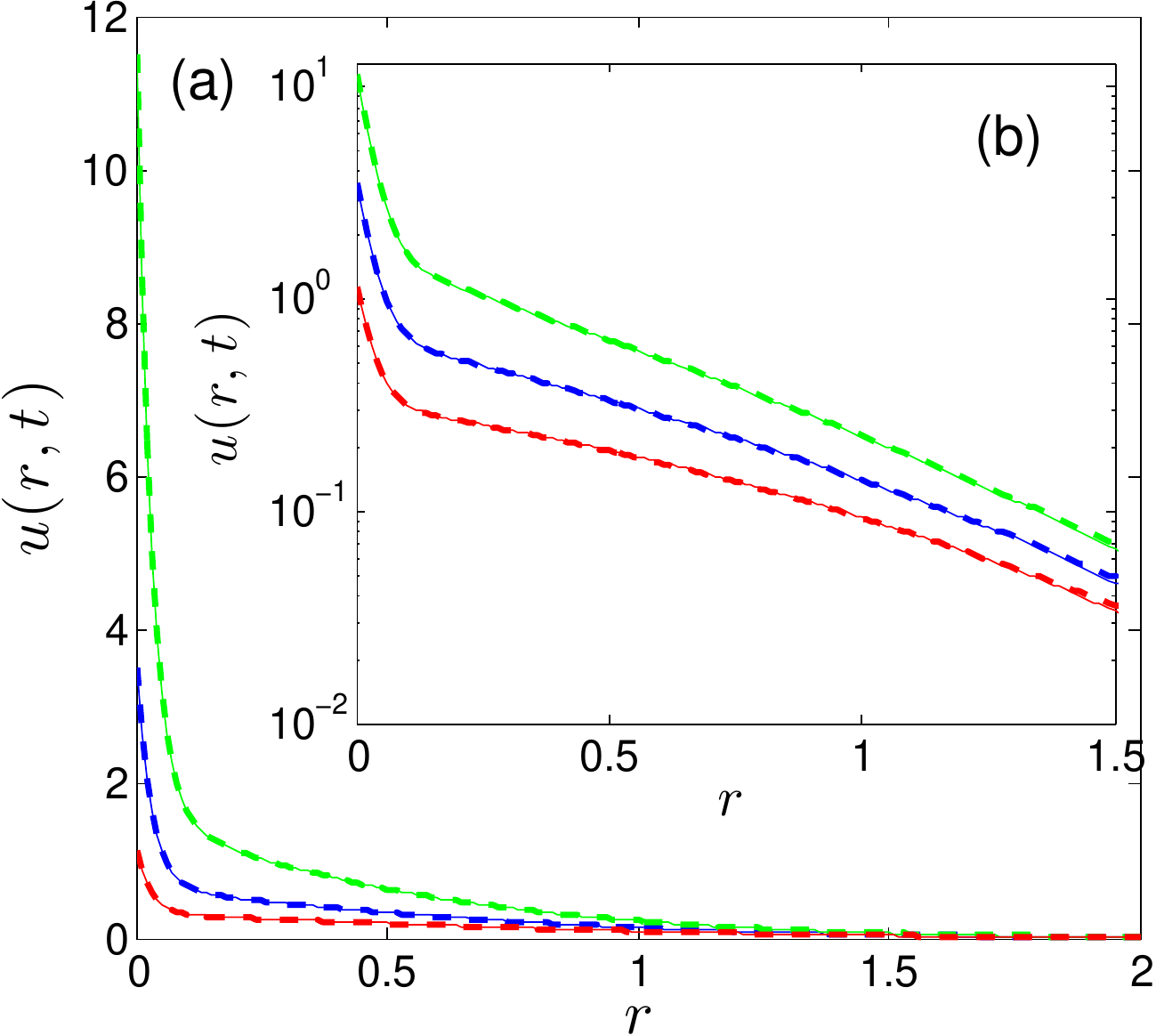}
\end{center}
\caption{(Color online) Evolution of the normalized and centered pair correlation function $u(r,t)$ for a one-dimensional domain (a), starting from a zero-reactivity equilibrium condition. Solid lines are the numerical solutions of the exact Eq.~\eqref{normalized_u_corr}, and dashed lines are the asymptotic solutions provided in Eqs.~\eqref{normalized_u_noncrit} and~\eqref{normalized_u_crit}. The pair correlation functions are displayed at time $t=3 \times 10^4$, with the following parameters: ${\mathcal N}_0=10^3$, $\eta = 8.333 \times 10^{-3}$, $\alpha \nu_m = 1.2$, $\lambda = 10^{-2}$, $\nu_n^{(2)} = 2$, $\nu_{nm} = 2.4 \times 10^{-2}$ and $\nu_m^{(2)} = 4\times 10^{-3}$. The presence of a peak close to the origin is the signature of spatial clustering. Red (lower) curves correspond to a supercritical reactor with $\rho = 5 \times 10^{-3}$, green (upper) curves correspond to a subcritical reactor with $\rho = -5 \times 10^{-3}$, and blue (central) curves correspond to an exactly critical reactor with $\rho = 0$. (b) An inset displays the same curves with a logarithmic scale on the ordinate axis.}
\label{fig7}
\end{figure}

\begin{figure}[t]
\begin{center}
\includegraphics[scale=0.6]{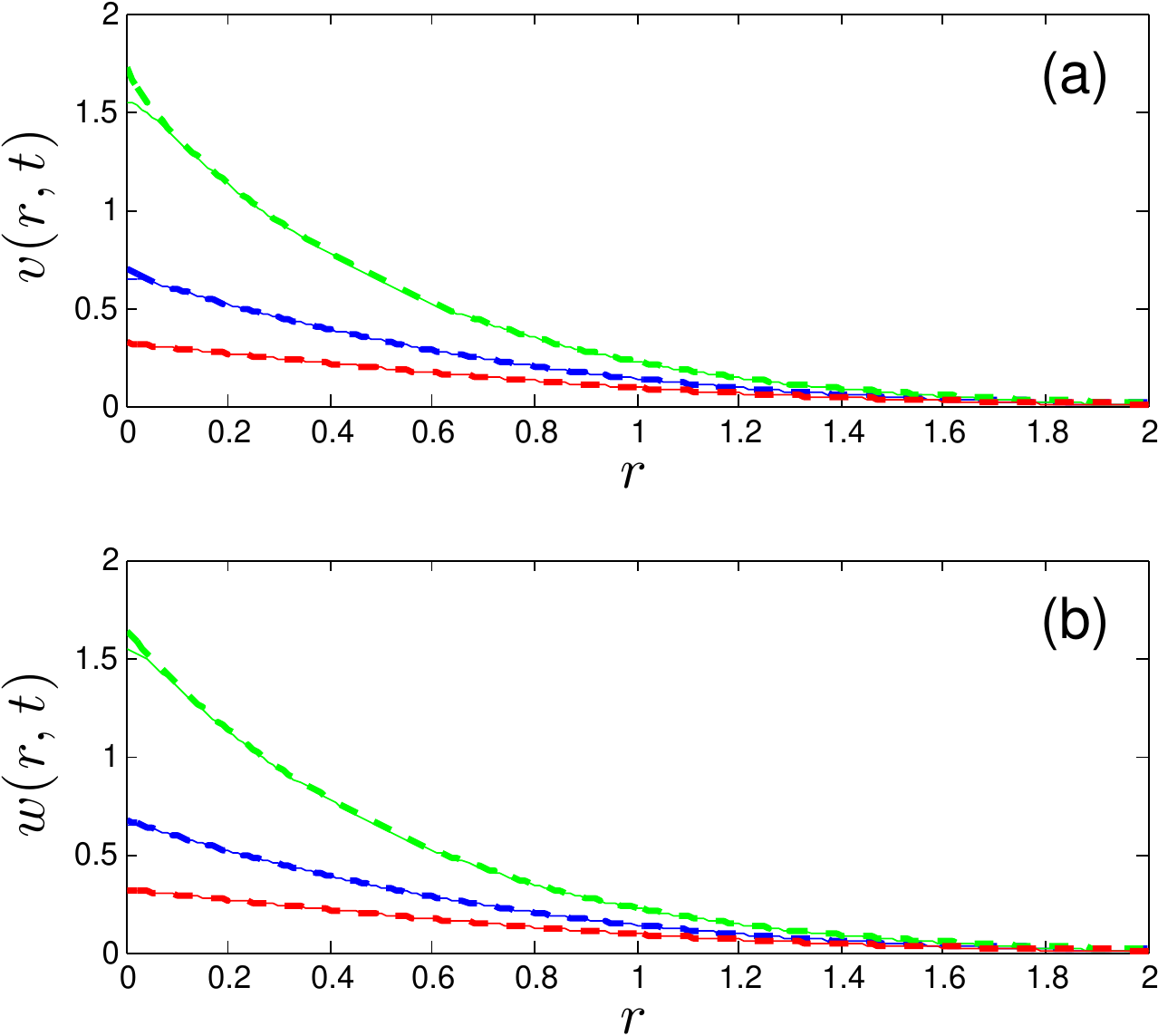}
\end{center}
\caption{(Color online) Evolution of the normalized and centered pair correlation function $v(r,t)$ (a) and $w(r,t)$ (b) for a one-dimensional domain, starting from a zero-reactivity equilibrium condition. Solid lines are the numerical solutions of the exact Eqs.~\eqref{normalized_v_corr} and~\eqref{normalized_w_corr}, respectively, and dashed lines are the asymptotic solutions provided in Eqs.~\eqref{normalized_v_noncrit} (Eq.~\eqref{normalized_v_crit} for the critical case) and~\eqref{normalized_w_noncrit} (Eq.~\eqref{normalized_w_crit} for the critical case), respectively. The pair correlation functions are displayed at time $t=3 \times 10^4$, with the following parameters: ${\mathcal N}_0=10^3$, $\eta = 8.333 \times 10^{-3}$, $\alpha \nu_m = 1.2$, $\lambda = 10^{-2}$, $\nu_n^{(2)} = 2$, $\nu_{nm} = 2.4 \times 10^{-2}$ and $\nu_m^{(2)} = 4\times 10^{-3}$. Red (lower) curves correspond to a supercritical reactor with $\rho = 5 \times 10^{-3}$, green (upper) curves correspond to a subcritical reactor with $\rho = -5 \times 10^{-3}$, and blue (central) curves correspond to an exactly critical reactor with $\rho = 0$.}
\label{fig8}
\end{figure}

The asymptotic expressions in Eqs.~\eqref{normalized_u_space_real},~\eqref{normalized_v_space_real} and~\eqref{normalized_w_space_real} are compared to the numerical solutions of the exact Eqs.~\eqref{normalized_u_corr},~\eqref{normalized_v_corr} and~\eqref{normalized_w_corr} in Figs.~\ref{fig7} and~\ref{fig8}. In the weak reactivity regime, the asymptotic solutions provide a remarkable approximation of the exact correlation function (small discrepancies are nonetheless visible for short times and distances, as expected). For long times, $u(r,t)$, $v(r,t)$ and $w(r,t)$ have the same asymptotic behaviour. When the reactor is subcritical, the average neutron and precursor densities decrease exponentially fast, and the spatial correlations diverge exponentially fast. On the contrary, when the reactor is supercritical the average neutron and precursor densities grow unbounded exponentially fast, and $u(r,t) \simeq (r,t) \simeq w(r,t)$ asymptotically flatten out as
\begin{align}
& u_\infty(r,t) \simeq \frac{\alpha \nu_m}{{\mathcal N}_0} \frac{\eta^2(A+ 2B +C)}{\omega(1+\eta)(1+\eta - \epsilon)} G(r,t),
\label{normalized_u_asy_space}
\end{align}
where we have used $F(r,t) \simeq G(r,t)/\omega$ for large times and $\omega>0$. The case of an exactly critical system leads to some simplifications: by taking the limit $\omega \to 0$ with $\epsilon =0$ and ${\mathcal D} = D /(1+\eta)$ we get
\begin{align}
u(r,t) &\simeq \frac{A' \left( \frac{\alpha \nu_m}{{\mathcal D}}\right)^{\frac{d+2}{4}}}{2 {\mathcal N}_0 (1+\eta)^2} \frac{K_{d/2-1}\left(r \sqrt{\alpha \nu_m/ {\mathcal D}}\right)}{(2 \pi)^{d/2} r^{d/2-1}} \nonumber \\
& +\frac{\alpha \nu_m}{{\mathcal N}_0} \frac{\eta (A'+ 2B' +C')}{(1+\eta)^2} \frac{\Gamma_{d/2-1}\left(\frac{r^2}{8 \eta {\mathcal D} t}\right)}{8 \pi ^{d/2} {\mathcal D} r^{d-2}} ,
\label{normalized_u_space_real_crit}
\end{align}
\begin{align}
v(r,t) &\simeq \frac{\eta (A'+ 2B') \left( \frac{\alpha \nu_m}{{\mathcal D}}\right)^{\frac{d+2}{4}}}{2 {\mathcal N}_0 (1+\eta)^2} \frac{K_{d/2-1}\left(r\sqrt{\alpha \nu_m/ {\mathcal D}}\right)}{(2 \pi)^{d/2} r^{d/2-1}} \nonumber \\
& +\frac{\alpha \nu_m}{{\mathcal N}_0} \frac{\eta (A'+ 2B' +C')}{(1+\eta)^2} \frac{\Gamma_{d/2-1}\left(\frac{r^2}{8 \eta {\mathcal D} t}\right)}{8 \pi ^{d/2} {\mathcal D} r^{d-2}},
\label{normalized_v_space_real_crit}
\end{align}
\begin{align}
w(r,t) &\simeq \frac{\alpha \nu_m}{{\mathcal N}_0} \frac{\eta (A'+ 2B' +C')}{(1+\eta)^2} \frac{\Gamma_{d/2-1}\left(\frac{r^2 }{8 \eta {\mathcal D} t}\right)}{8 \pi ^{d/2}{\mathcal D} r^{d-2}},
\label{normalized_w_space_real_crit}
\end{align}
$\Gamma_a(z)$ being the incomplete Gamma function~\cite{erdelyi}.

The expression of the neutron spatial correlation function $u(r,t)$ is to be compared with that of a reactor without precursors, for which we would have
\begin{equation}
u_p(r,t) = \frac{\alpha \nu_n^{(2)}}{{\mathcal N}_0} e^{-\omega_p t} \int_0^t dt' e^{\omega_p t'} \frac{\exp\left(-\frac{r^2}{8 D t'} \right) }{\left( 8 \pi D t'\right)^{d/2}}
\end{equation}
for $\epsilon \neq 0$, and
\begin{equation}
u_p(r,t) = \frac{\alpha \nu_n^{(2)} }{{\mathcal N}_0} \frac{\Gamma_{d/2-1}\Big( \frac{r^2}{8Dt}\Big)}{8\pi^{d/2}D r^{d-2} }
\end{equation}
for an exactly critical system~\cite{houchmandzadeh_pre_2009, zoia_pre_clustering}. By inspection, observing that the term $\nu_m (A'+2B'+C')$ is dominated by $\nu_n^{(2)}$ and that ${\cal D} \simeq D$, we finally have
\begin{equation}
u(r,t) \simeq \eta u_p(r, \eta t),
\end{equation}
in close analogy with the result for $u(t)$. Precursors are therefore extremely effective also in quenching the spatial clustering of the neutrons: in the presence of delayed neutrons, the spatial correlation function of the neutron population has a much slower evolution in time ($t \to \eta t$, similarly as for the case of the average density), and its amplitude is further rescaled by a factor $\eta$.

\section{Conclusions}
\label{conclusions}

Fission chains in nuclear reactors fluctuate in space and time due to the competition between neutron diffusion, births by fission and deaths by absorption. Because of the interplay between these three key mechanisms, the neutron population, although uniformly distributed in space at the initial time, may later display patchiness, a phenomenon which is known as spatial clustering. Close to the critical regime, the typical fluctuations may even become larger than the typical local neutron density, and thus make the average system behaviour meaningless. Nuclear reactors operated close to the critical point require that the population of prompt neutrons instantaneously emitted at fission events must be in equilibrium with the much smaller population of delayed neutrons, emitted after Poissonian times by the nuclear decay of the fissioned nuclei (the so-called precursors). The ratio $\eta = \lambda / (\alpha \nu_m)$ between the rate at which precursors disappear by giving rise to delayed neutrons ($\lambda$) and the rate at which precursors are created by fission events ($\alpha \nu_m$) plays a central role in determining the behaviour of the space-time fluctuations.

We have first explicitly derived the asymptotic expressions of the averages and the second moments of the total neutron and precursor populations, which are initially at equilibrium with a ratio $n_0/ m_0=\eta \ll 1$. In the presence of delayed neutrons induced by the decay of precursors, the average populations evolve in time much less rapidly (by a factor $\eta$) than for the case of a reactor with prompt neutrons alone. The normalized variance of the number of neutrons also evolves in time by a factor $\eta$ less rapidly, and its amplitude is further rescaled by a factor $\eta$ with respect to its purely prompt counterpart. This effect is due to the buffering effect of the precursor population.

Then, on the basis of these results, we have turned our attention to the case of the average spatial densities and the spatial correlation functions, for which we have derived the asymptotic expressions in the weak reactivity regime. Not entirely surprisingly, we have shown that the smoothing effect of precursors observed for the total populations carries over also to these physical observables. In particular, the spatial correlation function again evolves in time by a factor $\eta$ less rapidly, and its amplitude is further rescaled by a factor $\eta$ with respect to its purely prompt counterpart. This physically means that the equilibrium between neutrons and precursors is actually key in quenching the neutron fluctuations at the global scale as well as the spatial clustering at (and close to) the critical point.

\appendix

\section{Obtaining the moments from the master equation}
\label{master_moments}

We sketch here the derivation of the moment equations from the master equation, in order for the paper to be self-contained. A more thorough discussion can be found in~\cite{houchmandzadeh_pre_2002, houchmandzadeh_pre_2009}. Consider for instance a master equation in the form
\begin{align}
&\frac{\partial}{\partial t} {\cal P}_t(n) = W (n-1) {\cal P}_t(n-1) - W n {\cal P}_t(n),
\end{align}
where $W$ is a rate. Upon multiplying each term by a factor $n^m$ and summing over $n$, the left-hand-side immediately yields $\partial_t \langle n^m \rangle$. At the right-hand-side, a change of index $n \to n+1$ transforms
\begin{equation}
W \sum_n n^m (n-1){\cal P}_t(n-1)\to W \sum_n (n+1)^m n {\cal P}_t(n).
\end{equation}
Then, we get
\begin{align}
&\frac{\partial}{\partial t} \langle n^m \rangle  = W \langle n(n+1)^m \rangle - W \langle n^{m+1} \rangle.
\end{align}
Observe that $n(n+1)^m - n^{m+1} $ is a polynomial of order $m$. For instance, for the average we have $m=1$ and
\begin{align}
&\frac{\partial}{\partial t} \langle n \rangle  = W \langle n\rangle,
\end{align}
whereas for the second moment $m=2$ and
\begin{align}
&\frac{\partial}{\partial t} \langle n^2 \rangle  = 2 W \langle n^2\rangle + W \langle n\rangle.
\end{align}
The spatial behaviour can be obtained by following the same strategy. Observe that the creation and annihilation operators commute, i.e., $a_k a^\dag_k {\bf n} = a^\dag_k a_k {\bf n}$. Consider, e.g., a master equation in the form
\begin{align}
&\frac{\partial}{\partial t} {\cal P}_t({\bf n}) = \sum_{i} [W (n_i-1) {\cal P}_t(a_i {\bf n}) - W n_i {\cal P}_t({\bf n})].
\end{align}
Upon multiplication of each term by $n_k^m$ and summation over ${\bf n}$, the left-hand-side yields $\partial_t \langle n_k^m \rangle$. At the right-hand-side, the term $\sum_{{\bf n}} \sum_{i} n_k^m W (n_i-1) {\cal P}_t(a_i {\bf n})$ can be grouped with $- \sum_{{\bf n}} \sum_{i} n_k^m W n_i {\cal P}_t( {\bf n})$ by changing the summation variable ${\bf n} \to a^\dag_i{\bf n}$. This gives
\begin{align}
& W \sum_{{\bf n}} \sum_{i} [(n_k+\delta_{k,i})^m n_i - n_k^m n_i]{\cal P}_t( {\bf n}).
\end{align}
The only non-vanishing term of the sum over $i$ is then for $i=k$, which finally yields
\begin{align}
&\frac{\partial}{\partial t} \langle n^m_k \rangle = W \langle (n_k+1)^m n_k \rangle - W \langle n_k^{m+1} \rangle.
\end{align}
For instance, for the average we have
\begin{align}
&\frac{\partial}{\partial t} \langle n_k \rangle = W \langle n_k \rangle,
\end{align}
whereas for the second moment
\begin{align}
&\frac{\partial}{\partial t} \langle n^2_k \rangle = 2 W \langle n_k^2\rangle + W \langle n_k\rangle.
\end{align}

\section{Asymptotic analysis of the average total populations}
\label{app_mean}

Equations~\eqref{average_eq}, together with the initial conditions in Eqs.~\ref{ic_average}, can be solved exactly, and yield
\begin{align}
&\langle n(t) \rangle = n_0 \frac{(\rho-\Omega_2)e^{\Omega_1 t} + (\Omega_1-\rho)e^{\Omega_2 t}}{\Omega_1- \Omega_2} \\
&\langle m(t) \rangle = m_0 \frac{\Omega_2 e^{\Omega_1 t} - \Omega_1 e^{\Omega_2 t}}{\Omega_2- \Omega_1},
\label{average_eq_exact}
\end{align}
where the eigen-frequencies $\Omega_{1,2}$ are determined by the roots of the characteristic polynomial associated to~\eqref{average_eq}, namely,
\begin{align}
& \Omega_{1,2} = \frac{-\lambda + \rho -\alpha \nu_m \pm \sqrt{4\lambda \rho+ (\lambda - \rho +\alpha \nu_m)^2}}{2}.
\end{align}
Then, since $\Omega_1 \ge \Omega_2$ and $\Omega_2 < 0$, for long times $t \gg 1 / (\Omega_1 - \Omega_2)$ the moments asymptotically behave as
\begin{align}
&\langle n(t) \rangle \simeq n_0 \frac{\rho-\Omega_2}{\Omega_1- \Omega_2} e^{\Omega_1 t} \\
&\langle m(t) \rangle \simeq m_0 \frac{\Omega_2}{\Omega_2- \Omega_1} e^{\Omega_1 t}.
\label{average_eq_asy}
\end{align}
The sign of $\Omega_1$ depends on the reactivity $\rho$. The ratio between the two average populations asymptotically converges to a constant, namely,
\begin{align}
& \frac{\langle n(t) \rangle}{\langle m(t) \rangle} \simeq \eta \frac{\Omega_2-\rho}{\Omega_2}.
\end{align}
If the net reactivity is weak, expanding in small powers of $|\rho|$ yields the characteristic roots
\begin{align}
& \Omega_1 \simeq \frac{\eta}{1+\eta}\rho  \\
& \Omega_2 \simeq -\alpha \nu_m (1+\eta -\epsilon),
\end{align}
where we have introduced
\begin{align}
&\epsilon = \frac{\rho}{\alpha \nu_m} \frac{1}{1+\eta}.
\end{align}
For long times, the average densities will then exponentially grow or shrink with an asymptotic period
\begin{align}
& \omega = \frac{\eta}{1+\eta}\rho.
\end{align}

\section{Equations for the second moments}
\label{app_second_moments}

The equations for the second moments
\begin{align}
&\langle n^2(t) \rangle = \sum_{n,m} n^2 {\cal P}_t(n,m),\\
&\langle m^2(t) \rangle = \sum_{n,m} m^2 {\cal P}_t(n,m)
\end{align}
and for the cross-moment
\begin{equation}
\langle n(t) m(t) \rangle = \sum_{n,m} n m {\cal P}_t(n,m)
\end{equation}
are slightly cumbersome. After some manipulations (see Appendix~\ref{master_moments}), we get
\begin{align}
\frac{\partial}{\partial t}\langle n^2(t) \rangle  &= 2(\rho-\alpha \nu_m) \langle n^2(t) \rangle + 2 \lambda \langle n(t) m(t) \rangle \nonumber \\
& + (\alpha \nu_n^{(2)} + \alpha \nu_m-\rho )\langle n(t)\rangle + \lambda \langle m(t) \rangle,
\label{second_n_eq}
\end{align}
\begin{align}
\frac{\partial}{\partial t}\langle n(t) m(t) \rangle  &= (\rho -\alpha \nu_m -\lambda ) \langle n(t) m(t)\rangle \nonumber \\
& + \alpha \nu_m \langle n^2(t) \rangle + \lambda \langle m^2(t) \rangle + \alpha \nu_{nm} \langle n(t)\rangle  \nonumber \\
& - \alpha \nu_{m} \langle n(t)\rangle -\lambda \langle m(t)\rangle,
\label{second_nm_eq}
\end{align}
\begin{align}
\frac{\partial}{\partial t}\langle m^2(t) \rangle  & = 2 \alpha \nu_m \langle n(t) m(t) \rangle -2 \lambda \langle m^2(t)\rangle \nonumber \\
& + (\alpha \nu_m^{(2)} + \alpha \nu_m)\langle n(t) \rangle  +\lambda \langle m(t) \rangle ,
\label{second_m_eq}
\end{align}
where we have defined the factorial moments
\begin{align}
&\nu_n^{(2)} = \sum_{i,j} i(i-1) P_{i,j} \\
&\nu_m^{(2)} = \sum_{i,j} j(j-1 )P_{i,j}
\end{align}
and the cross-moment
\begin{equation}
\nu_{nm} = \sum_{i,j} ij P_{i,j}.
\end{equation}
These equations are to be solved together with the initial conditions $\langle n^2(0) \rangle = n^2_0$, $\langle n(0) m(0) \rangle  = n_0 m_0$ and $\langle m^2(0) \rangle  = m^2_0$. Similar results for the second moments have been previously obtained by several authors by following different strategies~\cite{williams, pazsit, natelson_theory, bell_delayed, sanchez}.

\section{Asymptotic analysis of the second moments of the total populations}
\label{app_uvw}

The system of differential equations~\eqref{normalized_u},~\eqref{normalized_v} and~\eqref{normalized_w} can be written in the compact form
\begin{equation}
\frac{1}{\alpha \nu_m}\frac{\partial}{\partial t} {\bf U}(t) = {\bf M} {\bf U}(t) + {\bf Q}\frac{e^{-\omega t}}{n_0},
\end{equation}
by setting ${\bf U}(t) = \left[ u(t),v(t),w(t) \right] ^T $,
\begin{equation}
{\bf M}=\left(
\begin{array}{ccc}
-2(1-\epsilon) & 2(1-\epsilon) & 0 \\
\eta & -1-\eta+\epsilon & 1-\epsilon\\
0 & 2\eta& -2\eta
\end{array}
\right)
\label{matrix_M}
\end{equation}
and ${\bf Q} = \left[ A,\eta B, \eta^2 C \right] ^T $, with
\begin{align}
& A = \frac{\nu_n^{(2)}}{\nu_m} \left( 1-\frac{\epsilon}{1+\eta}\right) + 2\left( 1-2\frac{\epsilon}{1+\eta}\right)\\
& B = \frac{\nu_{nm}}{\nu_m} -2\\
& C = \frac{\nu_m^{(2)}}{\nu_m}+2.
\end{align}
Then, by taking the Laplace transform of each term, we get the algebraic system
\begin{equation}
\left( {\bf M} - \frac{s}{\alpha \nu_m} {\bf I} \right)\tilde{{\bf U}}(s)   = - \frac{1}{n_0} \frac{1}{\omega + s}{\bf Q},
\label{laplace_0d}
\end{equation}
where $s$ denotes the Laplace variable and ${\bf I}$ the identity matrix. The asymptotic behaviour of the variances can be determined by solving the system~\eqref{laplace_0d} and expanding the transformed solutions $\tilde{{\bf U}}(s)= \left[ \tilde{u}(s),\tilde{v}(s),\tilde{w}(s) \right] ^T$ in dominant powers for small $s$. By retaining the leading order terms we get
\begin{align}
\tilde{u}(s) & \simeq \frac{\alpha \nu_m}{n_0} \frac{\eta^2(A+ 2B +C)}{(1+\eta)(1+\eta - \epsilon)} \tilde{F}(s) \nonumber \\
& +\frac{A(1-\epsilon)+\eta(3A+ 2B)}{2 n_0 (1+\eta)(1+\eta - \epsilon)} s \tilde{F}(s) ,
\label{normalized_u_noncrit_laplace}
\end{align}
\begin{align}
\tilde{v}(s) & \simeq \frac{\alpha \nu_m}{n_0} \frac{\eta^2(A+ 2B +C)}{(1+\eta)(1+\eta - \epsilon)} \tilde{F}(s) \nonumber \\
& +\frac{\eta (A+ 2B)}{2 n_0 (1+\eta)(1+\eta - \epsilon)}s \tilde{F}(s),
\label{normalized_v_noncrit_laplace}
\end{align}
\begin{align}
\tilde{w}(s) & \simeq\frac{\alpha \nu_m}{n_0} \frac{\eta^2(A+ 2B +C)}{(1+\eta)(1+\eta - \epsilon)} \tilde{F}(s),
\label{normalized_w_noncrit_laplace}
\end{align}
where we have defined
\begin{align}
\tilde{F}(s)=\frac{1}{s(\omega + s)} .
\end{align}
Then, by reverting to the real space we obtain
\begin{align}
u(t) & \simeq \frac{A(1-\epsilon)+\eta(3A+ 2B)}{2 n_0 (1+\eta)(1+\eta - \epsilon)}e^{-\omega t} \nonumber \\
& + \frac{\alpha \nu_m}{n_0} \frac{\eta^2(A+ 2B +C)}{(1+\eta)(1+\eta - \epsilon)} \frac{1-e^{-\omega t} }{\omega},
\end{align}
\begin{align}
v(t) & \simeq \frac{\eta (A+ 2B)}{2 n_0 (1+\eta)(1+\eta - \epsilon)} e^{-\omega t} \nonumber \\
& + \frac{\alpha \nu_m}{n_0} \frac{\eta^2(A+ 2B +C)}{(1+\eta)(1+\eta - \epsilon)} \frac{1-e^{-\omega t} }{\omega},
\end{align}
\begin{align}
& w(t) \simeq\frac{\alpha \nu_m}{n_0} \frac{\eta^2(A+ 2B +C)}{(1+\eta)(1+\eta - \epsilon)} \frac{1-e^{-\omega t} }{\omega}.
\end{align}

\section{Equations for the spatial correlations}
\label{app_spatial_second_moments}

By manipulating the master equation~\eqref{master_eq_space} (see Appendix~\ref{master_moments}), we obtain the evolution equations
\begin{align}
\frac{\partial}{\partial t} \langle n_k n_{k+j} \rangle & =2 (\rho - \alpha \nu_m) \langle n_k n_{k+j} \rangle + \gamma \langle n_k \Delta n_{k+j} \rangle \nonumber \\
& + \gamma \langle n_{k+j} \Delta n_k \rangle +\lambda (\langle n_{k+j} m_k \rangle + \langle n_{k} m_{k+j} \rangle) \nonumber \\
& + \delta_{j,0} ((\alpha \nu_n^{(2)} + \alpha \nu_m - \rho) \langle n_k \rangle + \lambda \langle m_k \rangle)\nonumber \\
& + \delta_{j,0} \gamma (\langle n_{k+1} \rangle + \langle n_{k-1} \rangle + 2 \langle n_{k} \rangle)\nonumber \\
& - \delta_{j,1} \gamma (\langle n_{k+1} \rangle + \langle n_{k} \rangle) \nonumber \\
&- \delta_{j,-1} \gamma (\langle n_{k} \rangle + \langle n_{k-1} \rangle),
\label{u_eq_space}
\end{align}
\begin{align}
\frac{\partial}{\partial t} \langle n_{k+j} m_k \rangle  & =(\rho - \alpha \nu_m -\lambda ) \langle n_{k+j} m_k \rangle + \gamma \langle m_k \Delta n_{k+j} \rangle  \nonumber \\
& +\alpha \nu_m \langle n_k n_{k+j} \rangle +\lambda \langle m_k m_{k+j} \rangle \nonumber \\
& + \delta_{j,0} (\alpha \nu_{nm} + \alpha \nu_m - \alpha \nu_m) \langle n_k \rangle \nonumber \\
& - \delta_{j,0} \lambda \langle m_k \rangle,
\label{v_eq_space}
\end{align}
\begin{align}
\frac{\partial}{\partial t} \langle m_k m_{k+j} \rangle  & = \alpha \nu_m (\langle n_{k+j} m_k \rangle + \langle n_{k} m_{k+j} \rangle) \nonumber \\
&  - \lambda \langle m_k m_{k+j} \rangle + \delta_{j,0} (\alpha \nu_m^{(2)} + \alpha \nu_m ) \langle n_k \rangle \nonumber \\
& + \delta_{j,0} \lambda \langle m_k \rangle.
\label{w_eq_space}
\end{align}

\section{Asymptotic analysis of the spatial correlations}
\label{app_uvw_spatial}

In the long time limit, ${\mathcal N}(t) \simeq {\mathcal N}_0 e^{\omega t}$ and $\chi_t \simeq \eta (1+\epsilon)$, so that  we can rewrite Eqs.~\eqref{normalized_u_corr},~\eqref{normalized_v_corr} and~\eqref{normalized_w_corr} as
\begin{equation}
\frac{1}{\alpha \nu_m}\frac{\partial}{\partial t} {\bf U}(r,t) =( {\bf M} + {\bf D} \nabla^2 ){\bf U}(r,t) + {\bf Q'} \frac{e^{-\omega t}}{{\mathcal N}_0} \delta(r),
\end{equation}
where we have defined the correlation vector ${\bf U}(r,t) = [u(r,t), v(r,t), w(r,t)]^T$, the rescaled diffusion matrix
\begin{equation}
{\bf D} = \frac{D}{\alpha \nu_m}\left(
\begin{array}{ccc}
2 & 0 & 0 \\
0 & 1 & 0 \\
0 & 0 & 0
\end{array}
\right).
\end{equation}
and ${\bf Q'} = \left[ A' ,\eta B', \eta^2 C' \right] ^T $, with $A' = \nu_n^{(2)}/\nu_m $, $B' = \nu_{nm}/ \nu_m$ and $C' = \nu_m^{(2)}/\nu_m$. The matrix ${\bf M}$ has been defined in Eq.~\eqref{matrix_M}. Observe that ${\bf M}$ and ${\bf D}$ do not commute.

By taking the Laplace and Fourier transforms, this system of partial differential equations reduces to an algebraic system for the transformed vector $\tilde{{\bf U}}(k,s) = [\tilde{u}(k,s), \tilde{v}(k,s), \tilde{w}(k,s)]^T$, namely,
\begin{equation}
\left( {\bf M} - k^2 {\bf D}  - \frac{s}{\alpha \nu_m} {\bf I} \right)\tilde{{\bf U}}(k,s)   = - \frac{1}{{\mathcal N}_0} \frac{1}{\omega + s}{\bf Q'},
\label{laplace_fourier}
\end{equation}
where $k$ denotes the Fourier variable. The asymptotic solution in time and space for the system is obtained by taking $s \to 0$ and $|k| \to 0$, respectively. By retaining the leading order terms for small $\epsilon$ and small $\eta$ we get
\begin{align}
\tilde{u}(k,s) &\simeq \frac{\alpha \nu_m}{{\mathcal N}_0} \frac{\eta^2(A'+ 2B' +C')}{(1+\eta)(1+\eta - \epsilon)} \tilde{F}(k,s)  \nonumber \\
& + \frac{\alpha \nu_m A'(1-\epsilon)}{2 {\mathcal N}_0 (1+\eta)(1+\eta - \epsilon)} \tilde{H}(k,s) 
\label{normalized_u_space_trans}
\end{align}
\begin{align}
\tilde{v}(k,s) & \simeq \frac{\alpha \nu_m}{{\mathcal N}_0} \frac{\eta^2(A'+ 2B' +C')}{(1+\eta)(1+\eta - \epsilon)} \tilde{F}(k,s) \nonumber \\
& + \frac{\alpha \nu_m \eta (A'+2B')}{2 {\mathcal N}_0(1+\eta)(1+\eta - \epsilon) } \tilde{H}(k,s) ,
\label{normalized_v_space_trans}
\end{align}
\begin{align}
\tilde{w}(k,s) & \simeq \frac{\alpha \nu_m}{{\mathcal N}_0} \frac{\eta^2(A'+ 2B' +C')}{(1+\eta)(1+\eta - \epsilon)} \tilde{F}(k,s),
\label{normalized_w_space_trans}
\end{align}
where we have set
\begin{align}
& \tilde{F}(k,s) = \frac{1}{(\omega+ s)\left( 2 \eta {\mathcal D} k^2 +s\right)},
\label{F_def}
\end{align}
with ${\mathcal D} = D / (1+\eta- \epsilon)$, and
\begin{align}
& \tilde{H}(k,s) = \frac{1}{(\omega+ s)\left( {\mathcal D}  k^2 + \alpha \nu_m\right)}.
\label{H_def}
\end{align}
For domains where spherical symmetry applies, the $d$-dimensional inverse Fourier transform $ f(r) = \mathcal{F}^{-1} [\tilde{f}(k)]$ may be expressed as~\cite{exp_flights}
\begin{align}
& f(r) =\frac{r^{1-d/2}}{(2\pi)^{d/2}}\int_0^\infty k^{d/2} J_{d/2-1}(kr) \tilde{f}(k) dk,
\end{align}
where $J_a(z)$ is the modified Bessel function of the first kind~\cite{erdelyi}. Now, observe that the function $F(r,t) = \mathcal{F}^{-1} [\mathcal{L}^{-1} [\tilde{F}(k,s)]] $ is given by
\begin{align}
& F(r,t) = e^{-\omega t} \int_0^t dt' e^{\omega t'} G(r,t'),
\end{align}
where $G(r,t)$ is the Gaussian function
\begin{align}
& G(r,t) = \frac{\exp\left(-\frac{r^2}{8 \eta {\mathcal D} t} \right) }{\left( 8 \pi \eta {\mathcal D} t\right)^{d/2}},
\end{align}
and the function $H(r,t) = \mathcal{F}^{-1} [\mathcal{L}^{-1} [\tilde{H}(k,s)]] $ is given by
\begin{align}
& H(r,t) =  \frac{e^{-\omega t}}{\alpha \nu_m} \left( \frac{\alpha \nu_m}{{\mathcal D}}\right)^{\frac{d+2}{4}} \frac{K_{d/2-1}\left(r \sqrt{\frac{\alpha \nu_m}{{\mathcal D}}}\right)}{(2 \pi)^{d/2} r^{d/2-1}}.
\end{align}
The solution ${\bf U}(r,t)$ in real space can be therefore expressed as:
\begin{align}
u(r,t) & \simeq \frac{\alpha \nu_m A'(1-\epsilon) }{2 {\mathcal N}_0 (1+\eta)(1+\eta - \epsilon)} H(r,t)\nonumber \\
& + \frac{\alpha \nu_m}{{\mathcal N}_0} \frac{\eta^2(A'+ 2B' +C')}{(1+\eta)(1+\eta - \epsilon)} F(r,t),
\end{align}
\begin{align}
v(r,t) & \simeq \frac{\alpha \nu_m \eta (A'+ 2B') }{2 {\mathcal N}_0 (1+\eta)(1+\eta - \epsilon)} H(r,t) \nonumber \\
& + \frac{\alpha \nu_m}{{\mathcal N}_0} \frac{\eta^2(A'+ 2B' +C')}{(1+\eta)(1+\eta - \epsilon)} F(r,t),
\end{align}
\begin{align}
w(r,t) & \simeq \frac{\alpha \nu_m}{{\mathcal N}_0} \frac{\eta^2(A'+ 2B' +C')}{(1+\eta)(1+\eta - \epsilon)} F(r,t).
\end{align}

\end{document}